# O/H-N/O: the curious case of NGC 4670[⋆]

Nimisha Kumari[1]†, Bethan L. James[2], Mike J. Irwin[1], Ricardo Amorín[3,4] &
Enrique Pérez-Montero[5]

[1]*Institute of Astronomy, University of Cambridge, CB3 0HA UK*
[2]*Space Telescope Science Institute, 3700 San Martin Dr, Baltimore, MD 21218*
[3]*Kavli Insititute for Cosmology, Cambridge UK*
[4]*Cavendish Laboratory, University of Cambridge, Cambridge CB3 0HE, UK*
[5]*Instituto de Astrofísica de Andalucía, CSIC, Apartado de correos 3004, E-18080 Granada, Spain*

**ABSTRACT**
We use integral field spectroscopic (IFS) observations from Gemini North Multi-Object Spectrograph (GMOS-N) of a group of four H II regions and the surrounding gas in the central region of the blue compact dwarf (BCD) galaxy NGC 4670. At spatial scales of ∼ 9 pc, we map the spatial distribution of a variety of physical properties of the ionised gas: internal dust attenuation, kinematics, stellar age, star-formation rate, emission line ratios and chemical abundances. The region of study is found to be photoionised. Using the robust direct $T_e$-method, we estimate metallicity, nitrogen-to-oxygen ratio and helium abundance of the four H II regions. The same parameters are also mapped for the entire region using the HII-CHI-mistry code. We find that log(N/O) is increased in the region where the Wolf-Rayet bump is detected. The region coincides with the continuum region, around which we detect a slight increase in He abundance. We estimate the number of WC4, WN2-4 and WN7-9 stars from the integrated spectrum of WR bump region. We study the relation between log(N/O) and 12 + log(O/H) using the spatially-resolved data of the FOV as well as the integrated data of the H II regions from ten BCDs. We find an unexpected negative trend between N/O and metallicity. Several scenarios are explored to explain this trend, including nitrogen enrichment, and variations in star formation efficiency via chemical evolution models.

## 1 INTRODUCTION

Chemical evolution of the Universe is one of the most explored topics in astrophysical research and is essential to unravel the secrets of cosmic origin. All chemical elements result from nucleosynthetic processes, which happened either after a few seconds of the Big Bang or in the first stars after the Dark Ages and then during the subsequent evolutionary stages of stars and are still happening today in the present-day galaxies. Consequently, this topic has been the subject of innumerable observational and theoretical studies, as we continue to investigate various aspects of chemical evolution (see for example Pagel 1997; Izotov & Thuan 1998; Matteucci 2003; Tremonti et al. 2004; Erb et al. 2006; Maiolino et al. 2008; Steigman 2007; Mannucci et al. 2010; Davé et al. 2012; Mollá et al. 2015).

The study of the relation between nitrogen-to-oxygen ratio and oxygen abundance has been a topical subject of investigation and debate in both observing (e.g. McCall et al. 1985; Thuan et al. 1995; Izotov et al. 2006; Amorín et al. 2010; Berg et al. 2012; James et al. 2015; Belfiore et al. 2017) and modelling (e.g. Edmunds 1990; Henry et al. 2000; Köppen & Hensler 2005; Mollá et al. 2006; Vincenzo et al. 2016) communities working on chemical evolution. Nitrogen is of special interest as its origin is both primary and secondary. It may be produced from stars whose gas mixture contain only H and He (primary origin), and also from stars whose initial gas mixture contain metals (secondary origin). Thus the production of nitrogen may or may not depend on the initial metallicity of the gas. The relative abundance of nitrogen and oxygen is regulated by various factors such as star-formation history, presence of low and high mass stars, local chemical pollution possibly due to supernovae or wolf-rayet (WR) stars, and the flow of gas in, out and within the galaxies (Edmunds 1990; Henry et al. 2000; Köppen & Hensler 2005; Mollá et al. 2006; Vincenzo et al. 2016). Hence, mapping the

---

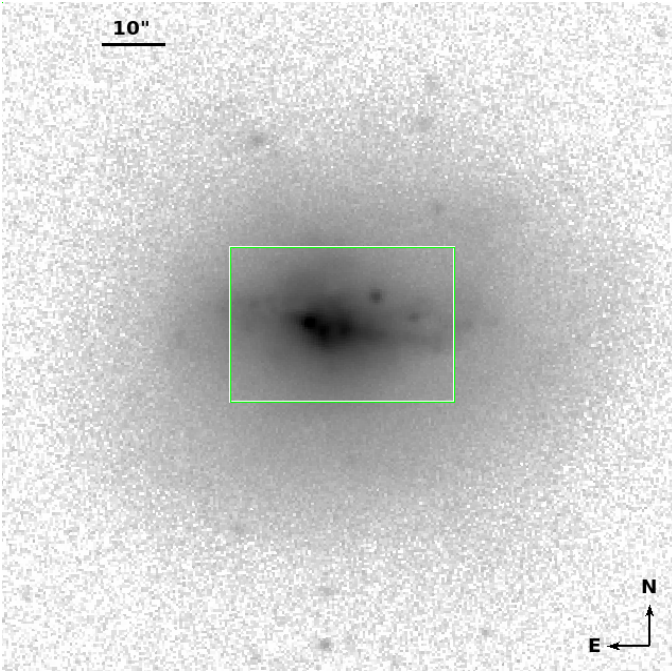

**Figure 1.** SDSS r-band image, showing the bright central region and an elliptical halo of older stellar population in NGC 4670. The green rectangular box shows region covered by the HST image shown in Figure 2.

distribution of physical properties of the ionised gas within galaxies is essential in understanding not only the nucleosynthetic origin of nitrogen but also the chemical enrichment and recycling processes.

Chemical abundances of galaxies can be robustly estimated and mapped by using the direct $T_e$-method, which requires the detection of weak auroral lines (e.g. [O III] $\lambda$ 4363, [N II] $\lambda$ 5755). In the absence of such detections, indirect methods are used for estimating the chemical abundances, which involve the use of strong emission lines. These indirect methods may be either the well-established calibrations involving the emission line ratios (e.g. Pettini & Pagel 2004; Maiolino et al. 2008; Pérez-Montero & Contini 2009; Dopita et al. 2016; Curti et al. 2017), or using the emission line fluxes in the photoionisation models of the ionised nebulae, such as CLOUDY (Ferland et al. 2013), MAPPINGS (Sutherland & Dopita 1993). Some examples of the codes which use such models to calculate abundances are HII-CHI-mistry (Pérez-Montero 2014), IZI (Blanc et al. 2015) and BOND (Vale Asari et al. 2016).

Blue compact dwarf galaxies (BCD, Searle & Sargent 1972; Thuan & Martin 1981) in the local Universe are ideal laboratories for mapping chemical abundance as they host luminous H II regions which emit in the visible range, hence providing a plethora of the emission lines required for chemical abundance analysis. Moreover, BCDs are low-metallicity (1/50–1/3 $Z_\odot$), starbursting dwarf galaxies in the nearby Universe, whose properties (for example, metallicity, compactness, specific star-formation rate, gas fraction) resemble those which are observed in primeval galaxies at high redshift (see Kunth & Östlin 2000, for a review). Their proximity enables detailed in-depth analyses of a variety of physical properties for both the young (Hunter & Thronson 1995; Papaderos et al. 1998; Thuan et al. 1999) and more evolved stellar components (Papaderos et al. 1996; Gil de Paz et al. 2003; Cairós et al. 2007; Amorín et al. 2009). However most of these analyses are based on either long-slit and/or photometric observations (e.g. Izotov & Thuan 1999; Cairós et al. 2001a,b; Hägele et al. 2012), which do not allow us to simultaneously map the chemical abundances and other physical properties within the BCDs. As such, any information on the spatial correlation between different physical properties and chemical abundance patterns are lost.

**Table 1.** General Properties of NGC 4670

| Parameter | NGC 4670 |
|---|---|
| Other designation | UGC 07930, Haro 9, Arp 163 |
| Galaxy Type | BCD, WR |
| R.A. (J2000.0) | 12h45m17.1s |
| DEC (J2000.0) | +27d07m31s |
| Redshift (z)$^a$ | 0.003566 ± 0.000013 |
| Distance (Mpc)$^a$ | 18.6 |
| inclination (°)$^b$ | 28 |
| Helio. Radial Velocity(km s$^{-1}$)$^a$ | 1069 ± 4 |
| E(B-V)$^c$ | 0.0128 ± 0.0003 |
| M$_B^b$ | −18.6 |
| U-B$^b$ | −0.49 |
| 12 + log(O/H)$^d$ | 8.30 |
| M$_*$ (M$_\odot$)$^e$ | $10^{8.78^{+0.2}_{-0.17}}$ |

$^a$ Taken from NED
$^b$ Hunter et al. (1996)
$^c$ Foreground galactic extinction (Schlafly & Finkbeiner 2011)
$^d$ Hirashita et al. (2002)
$^e$ SDSS

With the advent of integral field spectroscopy (IFS), the study of BCDs have been revolutionised (e.g. James et al. 2009, 2010, 2013a,b; Westmoquette et al. 2013; Lagos et al. 2012, 2014, 2016) as IFS has enabled to spatially-resolve the distribution of the physical properties of the ISM within the BCDs. This has allowed statistical analysis of these distributions to explore the spatial uniformity or homogeneity of such properties within BCDs (Pérez-Montero et al. 2011, 2013; Kehrig et al. 2013, 2016). IFS also allows us to analyse the spatial correlation between different properties, for example using IFS, Kehrig et al. (2008) detected both a WR population and an excess N/O across the BCD IIZw70; López-Sánchez et al. (2011) detected WR features and He II $\lambda$4686 emission line at the same location in the BCD Ic10, and Kumari et al. (2017) reported signatures of shock ionisation in the spatially-resolved emission line ratio diagrams and the velocity structure of the ionised gas of the central H II region of the BCD NGC 4449. Hence IFS studies are essential in understanding the cause and effect of various physical processes taking place in the ISM of the BCDs.

This paper is the second in a series of IFS analyses of star-forming regions in BCDs (see Kumari et al. 2017), where we aim to gain a deeper insight into the physical properties of these systems, by answering the following related questions: (1) How are the nitrogen-to-oxygen ratio (N/O) and oxygen abundance (O/H) distributed in the region of study? (2) Which physical mechanism is primarily responsible for the ionisation of gas? (3) What is the age of the stellar population currently ionising the gas in the target region of study? In this paper, we have targeted the central region of NGC 4670 (Figure 1), a BCD (Gil de Paz et al. 2003) which has appeared in many studies comprising large sample of star-forming galaxies (e.g. Hunter et al. 1982; Hunter 1982; Kinney et al. 1993; Moustakas & Kennicutt 2006; Brauher et al. 2008; Haynes et al. 2011; James et al. 2014).

NGC 4670 is an ideal target for addressing these questions and specifically for studying the relation between N/O and O/H. Beside being classified as a BCD, it has also been recognised as a WR galaxy (Mas-Hesse & Kunth 1999). As such, we expect local chemical pollution resulting from the winds of the WR stars, which may lead to chemical enrichment of the ISM and an enhanced N/O. Being a BCD, it is expected to have a low-metallicity and high star-formation rate. In fact, neutral hydrogen observations of NGC 4670 show a high concentration of the gas in the centre of the galaxy (Hunter et al. 1996), which may also increase the star-formation. A multi-wavelength analysis of the galaxy by Huchra et al. (1983) indicates the presence of several hundred O stars in giant H II region complexes, which indicate that the galaxy hosts young stellar populations. It is therefore possible that the metallicity, star-formation or stellar properties might be spatially-correlated with the WR stellar population or N/O. Moreover, this galaxy is relatively close (∼ 18.6 Mpc), as such the high spatial sampling (0.1″) provided by the integral field unit (IFU) on the Gemini Multi Object Spectrograph North (GMOS-N) has allowed us to study the spatial correlation between these properties at a spatial scale of 9 pc. A previous IFS study of the entire galaxy was performed using Visible Integral Field Replicable Unit Spectrograph (VIRUS-P), but had a spatial sampling of 4.2″ or 350 pc Cairós et al. (2012). Our study presents for the first time the IFS observation of NGC 4670, at a very fine spatial sampling of 0.1″ hence allowing a detailed analysis of spatial properties. Moreover the IFS data has enabled us to identify four new luminous H II regions in this galaxy. General properties of NGC 4670 are given in Table 1.

Table 2. GMOS-N IFU observing log for NGC 4670

| Grating | Central wavelength (Å) | Wavelength Range (Å) | Exposure Time (s) | Average Airmass | Standard Star |
|---|---|---|---|---|---|
| B600+_G5307 | 4650 | 3196 – 6067 | 2×1550 | 1.16, 1.24 | Hz44 |
| B600+_G5307 | 4700 | 3250 – 6118 | 3×1550 | 1.015, 1.034, 1.066 | Wolf1346 |
| R600+_G5304 | 6900 | 5345 – 8261 | 2×1400 | 1.013, 1.008 | Wolf1346 |
| R600+_G5304 | 6950 | 5397 – 8314 | 2×1400 | 1.12, 1.038 | Wolf1346 |

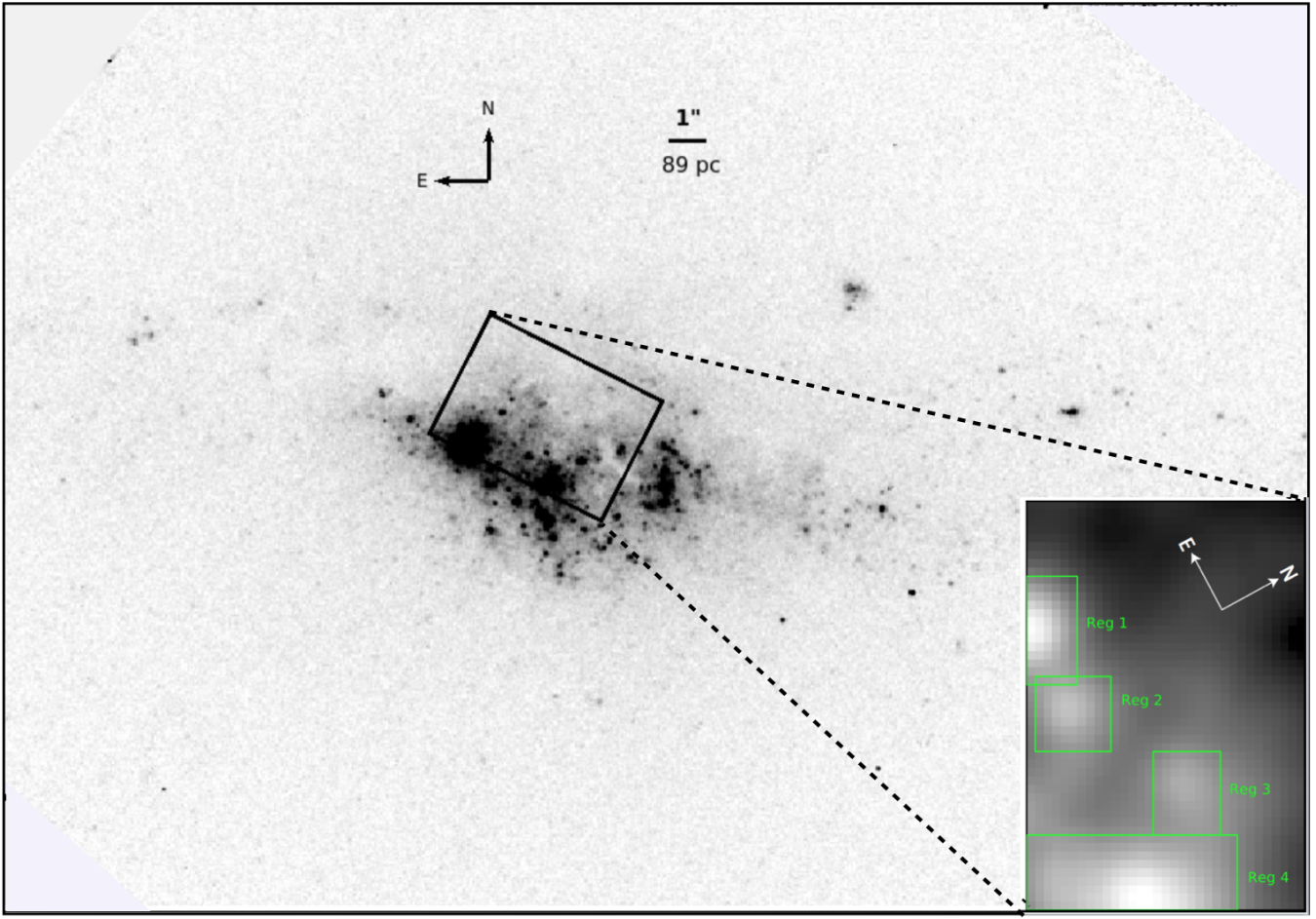

**Figure 2.** Left panel: HST image of NGC 4670 taken in the filter F439W. The black rectangular box in the centre represents the GMOS aperture (3.5″ × 5″). The HST image has a spatial scale of 0.05″ pixel$^{-1}$. North and East on the image is shown by the compass on the top-left of the figure. Right panel: H$\alpha$ map of the FOV obtained from the GMOS-IFU shows the four H II regions (Reg 1, Reg 2, Reg 3, Reg 4) in green rectangular boxes. The compass on this panel shows North and East on our FOV.

The paper is organised as follows: Section 2 presents the ob-servation and data reduction. Section 3 presents the preliminary procedures required for further data analysis and the main results, which include the gas kinematics, chemical abundances and stellar properties. In section 4, we explore the relation between the nitrogen-to-oxygen ratio and oxygen abundance at the spatially-resolved scale, and for the H II regions in this BCD. We also do a comparative analysis with a sample of H II regions within nine more BCDs compiled from the literature, a sample of green peas and also a large sample of star-forming galaxies from the Sloan Digital Sky Survey (SDSS). We explore the observed trend in the relation with the chemical evolution models. We finally summarize our results in section 5.

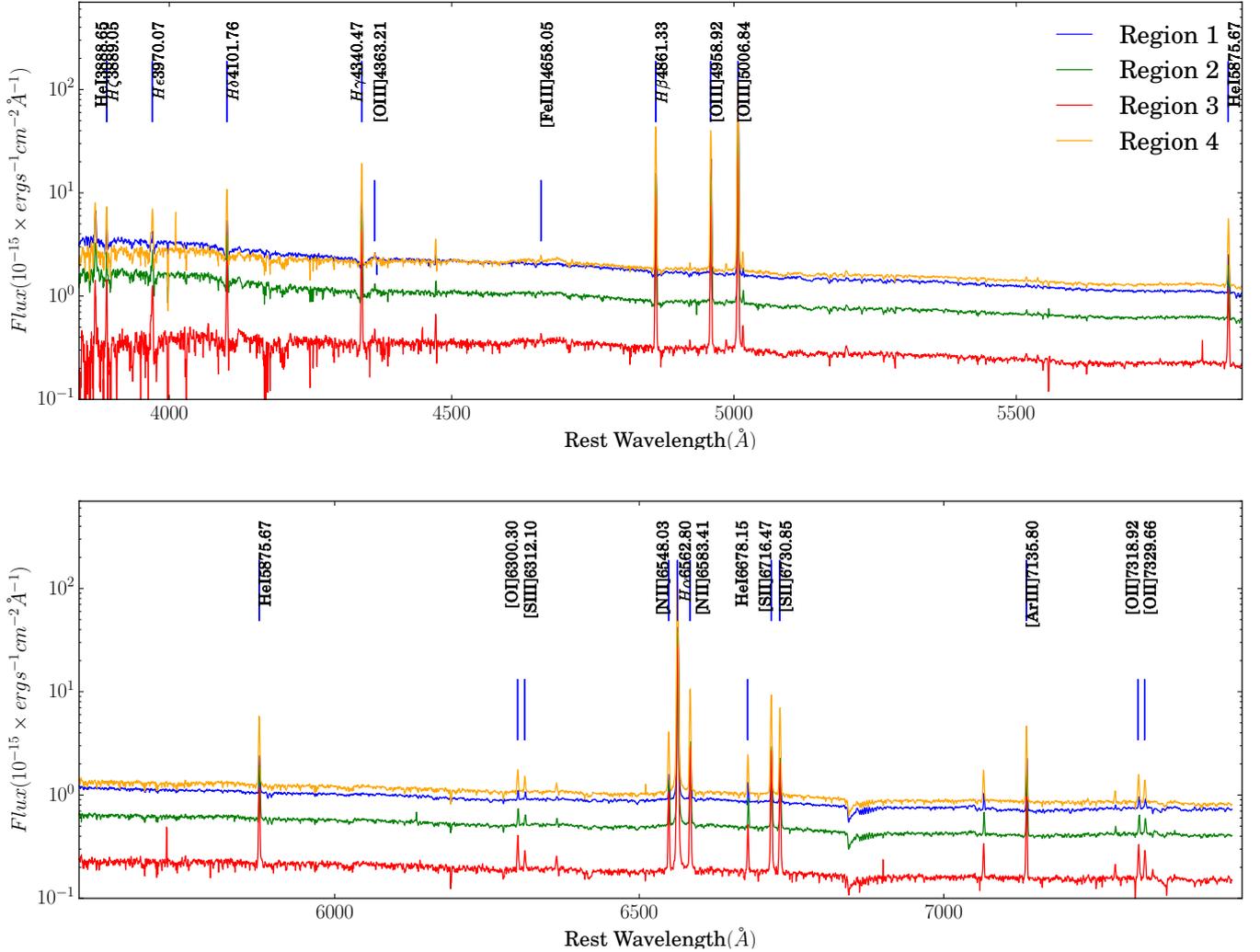

**Figure 3.** GMOS-IFU integrated spectra of individual H II regions in the FOV in the blue (upper panel) and red (lower panel) settings. The spectra of the four H II regions are colour-coded as follows: region 1: blue, region 2: green, region 3: red, region 4: orange. The blue ends of the spectra are noisier due to the low sensitivity of GMOS-IFU, which makes it difficult to show [O II] $\lambda$3727,3729 line in log scale. We show the [O II] $\lambda$3727,3729 line for all regions in Figure 4, along with the spectral line fitting.

## 2 OBSERVATION & DATA REDUCTION

The target region of NGC 4670 was observed with the Gemini Multi-Object Spectrograph (GMOS; Hook et al. (2004)) and IFU (GMOS-N IFU; Allington-Smith et al. 2002) at Gemini-North telescope in Hawaii, in one-slit queue-mode. This observation mode provides a field-of-view (FOV) of $3.5'' \times 5''$ sampled by 750 hexagonal lenslets of projected diameter of $0.2''$, of which 250 lenslets are dedicated to sky background determination. Table 2 presents information from the data observing log. Observations were carried out in four different settings using grating B600+_G5307 (B600) and grating R600+_G5304 (R600) covering the blue and red regions of the optical spectrum respectively. To avoid the problems in wavelength coverage due to two chip gaps between the three detectors of GMOS-N IFU, two sets of observations were taken with spectral dithering of 50 Å. For each of the four settings, a set of standard observations of GCAL flats, CuAr lamp for wavelength calibration and standard star for flux calibration were taken.

The basic steps of data reduction including bias subtraction, flat-field-correction, wavelength calibration and sky subtraction and differential atmospheric correction, were carried out using the stan-dard GEMINI reduction pipeline written in Image Reduction and Analysis Facility (IRAF)[1]. However the standard pipeline does not provide satisfactory results for some procedures and we therefore had to develop and implement our own codes. For example, wavelength calibration of the observations in one of the red-settings and the flux calibration of the observations in one of the blue-settings did not agree with the observations in the other three settings. We corrected the offset in wavelength calibration by comparing with the redshift obtained from the blue setting. Similarly, we statistically determined the scaling factor in the flux of the spectra in the blue setting. More information about these and other corrections procedures can be found in Kumari et al. (2017). We used the routine *gfcube* available in Gemini's IRAF reduction package to convert the spectra in each setting into three-dimensional data cubes, where we chose a spatial sampling of 0.1″ which was adequate to preserve the hexagonal sampling of GMOS-IFU lenslets. We corrected the spatial offset and spectral dithering between the observations of the same grating while combining the cubes obtained from that grating. The FOV covered by the two gratings (B600 and R600) showed a spatial offset of 0.1″ and 0.2″ in the x- and y-axes (with Figure 2 as reference). We produced cubes and row-stacked spectra of the overlapping regions of the FOVs covered by the two gratings, which we used for further analysis. For both red and blue setting, we fitted a Gaussian profile to several emission lines of the extracted row stacked spectra of the arc lamp, and found the value of instrumental broadening (Full Width Half Maximum, FWHM) to be ∼ 1.7 Å.

## 3 RESULTS

### 3.1 Observed and Intrinsic Fluxes

#### 3.1.1 Flux measurement

Figure 2 (left panel) shows the Hubble Space Telescope (HST) image of NGC 4670 taken in the filter F435W. The black rectangular box represents the GMOS aperture (3.5″ × 5″). The lower right panel presents the distribution of H$\alpha$ emission line (obtained from GMOS) across the FOV, and clearly shows four regions (green rectangular boxes labelled as "Reg 1", "Reg 2", "Reg 3" and "Reg 4") of current/increased star-formation activity. These regions have been selected by visually inspecting the H$\alpha$ emission line map, and roughly identifying isophotal regions. The present analysis includes the spatially-resolved and integrated properties of these four H II regions, which are referred to as Regions 1, 2, 3 and 4 in the following. Note here that all analysis related to Regions 1 & 4 should be treated with more caution since the GMOS-FOV does not cover these regions completely. In Figure 3, we show the GMOS-IFU integrated spectra of these four regions in the blue and red parts of the optical spectrum. The principal emission lines are over-plotted at their rest wavelengths in air.

We measure the emission line fluxes for all the main recombination and collisionally excited lines within the spectra by fitting Gaussian profiles to emission lines after subtracting the continuum and absorption features in recombination lines in the spectral region of interest. A single Gaussian profile was used to fit each emission line. Figure 4 shows the Gaussian fits for the [O II] doublet, where the peaks of the two emission lines [O II]$\lambda\lambda$3727,3729 could be

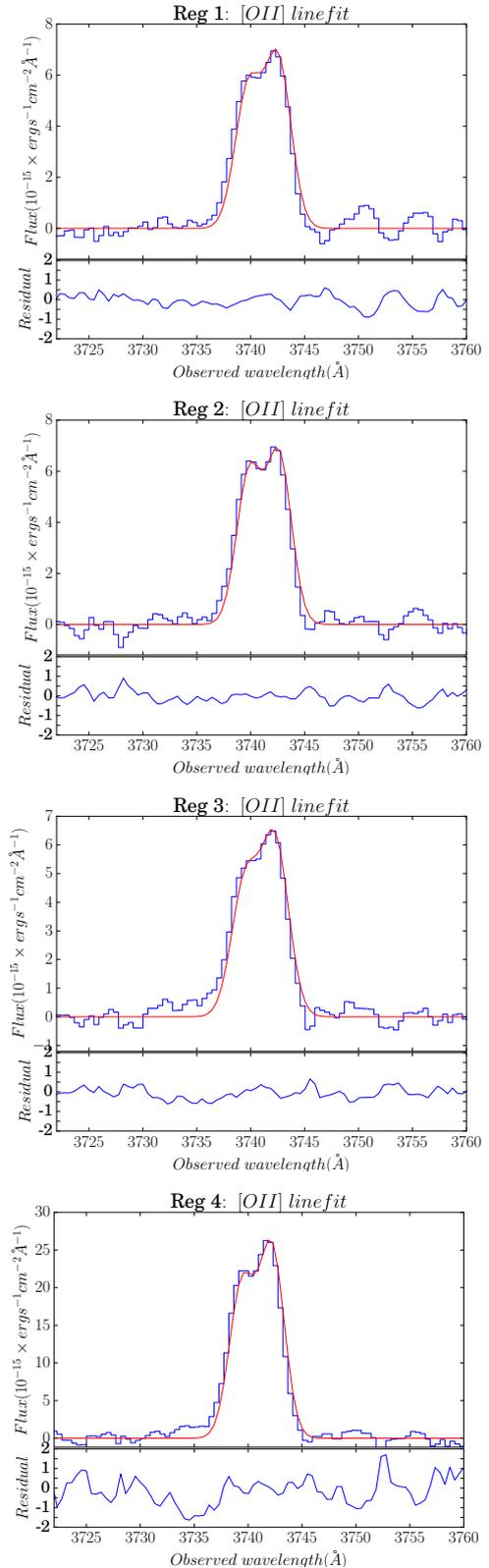

**Figure 4.** In the four panels corresponding to four H II regions, the upper panel shows the continuum-subtracted [O II]$\lambda\lambda$3727,3729 line (blue curve) detected in each region along with the 'Gaussian' fit (red curve) to extract flux, and the lower panels show the residuals normalised to be in $\sigma$-noise units.

---

[1] IRAF is distributed by the National Optical Astronomy Observatory, which is operated by the Association of Universities for Research in Astronomy (AURA) under a cooperative agreement with the National Science Foundation.

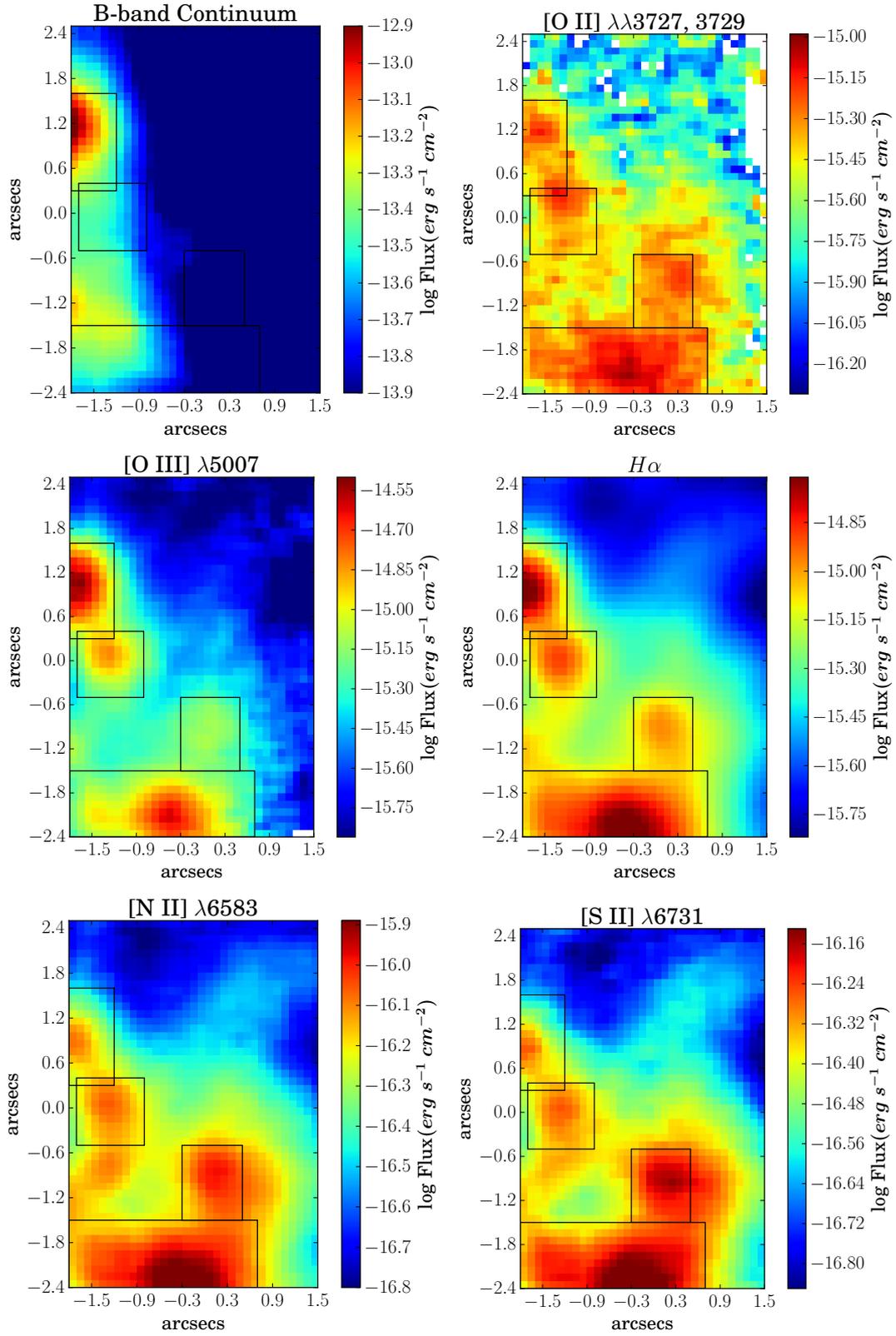

**Figure 5.** Observed B-band continuum map and the emission line flux maps ( [O II]$\lambda\lambda$3727,3729, [O III]$\lambda$5007, H$\alpha$, [N II]$\lambda$6584 and [S II]$\lambda$6717) of NGC 4670. The four black rectangular boxes denote the location of the four H II regions. White spaxels correspond to the spaxels in which emission line fluxes had S/N < 3.

### 3.1.2 Dust attenuation

We estimate E(B-V) by using the relationship between the nebular emission line colour excess and the Balmer decrement given by:

$$E(B-V) = \frac{E(H\beta - H\alpha)}{k(\lambda_{H\beta}) - k(\lambda_{H\alpha})}$$
$$= \frac{2.5}{k(\lambda_{H\beta}) - k(\lambda_{H\alpha})} log_{10}\left[\frac{(H\alpha/H\beta)_{obs}}{(H\alpha/H\beta)_{theo}}\right] \quad (1)$$

where $k(\lambda_{H\beta})$ and $k(\lambda_{H\alpha})$ are the values from the LMC (Large Magellanic Cloud) attenuation curve (Fitzpatrick 1999)[2] evaluated at the wavelengths $H\beta$ and $H\alpha$ respectively, $(H\alpha/H\beta)_{obs}$ and $(H\alpha/H\beta)_{theo}$ denote the observed and theoretical $H\alpha/H\beta$ line ratios respectively. We chose the LMC attenuation curve because the metallicity of NGC 4670 is reported to be 12+log(O/H) = 8.30 (Hirashita et al. 2002), which is close to that of LMC, 8.35±0.06 (Russell & Dopita 1992).

Following the above procedure, we found negative values of E(B-V) for some spaxels in random regions of the FOV, which we forced to the "Galactic foreground" E(B-V) value (= 0.0128). The negative values of E(B-V) could be due to dominance of shot noise in the low-extinction regions (Hong et al. 2013; Kumari et al. 2017). Figure 6 shows the map of E(B-V), which varies from 0.012–0.60 mag across the FOV. We find that the four H II regions have relatively lesser dust attenuation than the rest of the regions in the FOV. The E(B-V) map appears to be similar to the continuum map (Figure 5, upper right panel), i.e. the region to north-west of the group of H II regions with weak continuum is more extincted. To investigate this further, we compared the continuum maps of NGC 4670 in the B-band and R-band, our color (B-R) map resembles the E(B-V) map. The Balmer decrements ($H\alpha/H\beta$) observed for the four H II regions are in agreement with the results of Huchra et al. (1983), who report it to be approximately 3 from the broadband photometric data of NGC 4670. Our results are consistent with works of Huchra et al. (1983), whose optical data indicate little internal extinction.

Using E(B-V) obtained above, we calculate the extinction in magnitudes for the emission line fluxes given by $A_\lambda(mag) = k(\lambda)E(B-V)$ and finally calculate the intrinsic flux maps using the following equation:

$$F_{int}(\lambda) = F_{obs}(\lambda) \times 10^{0.4 A_\lambda} \quad (2)$$

The intrinsic emission line fluxes calculated from the integrated spectra of the four H II regions are presented in Table 3.

## 3.2 Gas Kinematics

Figure 7 shows the maps of radial velocity and velocity dispersion (FWHM) of the H$\alpha$ emission line, obtained from the centroid and width of the Gaussian fit to the emission line. We correct the radial velocity map for the barycentric correction (= −13.74 km s$^{-1}$) and systemic velocity of 1069 km s$^{-1}$ of NGC 4670 (Wolfinger et al. 2013). We also correct the FWHM maps for the instrumental broadening of 1.7 Å of GMOS-IFU.

The radial velocity map (Figure 7, left panel) shows that the ionised gas is slowly rotating about an axis of rotation going diagonally (NE-SW) through the FOV. The radial velocity varies between ∼ −10 to 30 km s$^{-1}$. The gas is redshifted in H II regions 1 and 2 hardly resolved for the four star-forming regions. In the spaxels where the peaks of the two emission lines in the [O II] doublet could not be resolved at all, we fitted a single Gaussian. While fitting Gaussian to the emission lines, we gave equal weight to flux in each spectral pixel since the dispersion in the continuum flux is found to be constant in the spectral region of interest both in the calibrated and uncalibrated spectra. This shows that the uncertainty of flux-determination within a spectral window is constant. Our error estimates on fluxes are the errors obtained while fitting Gaussians to the emission lines. The uncertainty related to the level of continuum is very small, hence the error on the flux is dominated by the error on the Gaussian fitting. The error estimates are consistent with those estimated from a Monte Carlo simulation. The fitting errors have been propagated to the other quantities using Monte Carlo simulations in subsequent analysis.

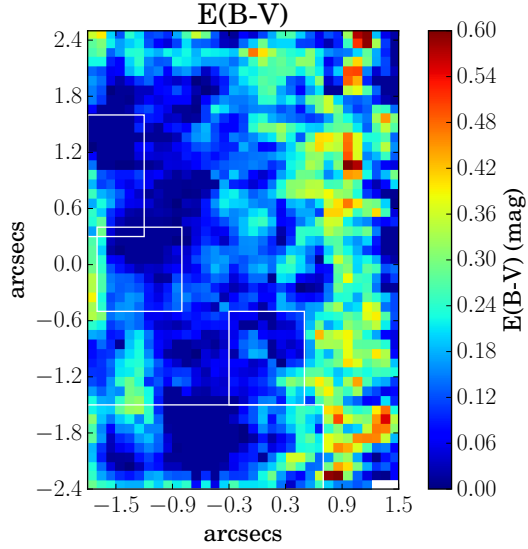

**Figure 6.** E(B-V) map created assuming the LMC extinction curve. Spaxels with E(B-V) < 0 are set to E(B-V) of the Galactic foreground. The four white rectangular boxes denote the location of the four H II regions. White spaxels correspond to the spaxels in which emission line fluxes had S/N < 3.

Figure 5 shows the observed flux maps of the GMOS FOV of B-band continuum, [O II]$\lambda\lambda$3727,3729, [O II]$\lambda$5007, H$\alpha$, [N II]$\lambda$6584 and [S II]$\lambda$6717. White spaxels in all maps correspond to the spaxels in which emission lines have S/N < 3. The B-band continuum map is obtained by integrating the blue cube in the wavelength range of 3980–4920 Å(in the rest-frame). The spatial profile of the continuum map remains same irrespective of the masking of emission lines. Table 3 presents the observed fluxes for the main emission lines used in the present analysis obtained from the integrated spectra of the four H II regions (Figure 3; "Reg 1", "Reg 2", "Reg 3" and "Reg 4" shown in Figure 2).

---

[2] The choice of attenuation curves is of little importance here as the attenuation curves of LMC, SMC, starburst or the MW have similar values in the optical range.

**Table 3.** Emission line measurements (relative to Hβ = 100) for the integrated spectra of the four HII regions shown in Figure 2. Line fluxes ($F_\lambda$) are extinction corrected using E(B-V) to calculate $I_\lambda$ for each of the individual H II regions.

| Line | $\lambda_{air}$ | $F1_\lambda$ | $I1_\lambda$ | $F2_\lambda$ | $I2_\lambda$ | $F3_\lambda$ | $I3_\lambda$ | $F4_\lambda$ | $I4_\lambda$ |
|---|---|---|---|---|---|---|---|---|---|
| [OII] | 3726.03 | 127.88 ± 3.63 | 135.57 ± 6.02 | 147.14 ± 4.43 | 154.48 ± 7.10 | 178.38 ± 6.28 | 197.62 ± 8.68 | 153.60 ± 3.22 | 169.56 ± 5.85 |
| Hγ | 4340.47 | 44.82 ± 0.80 | 45.97 ± 1.66 | 43.88 ± 0.56 | 44.35 ± 1.54 | 43.08 ± 0.52 | 45.45 ± 1.22 | 42.76 ± 0.35 | 44.63 ± 1.18 |
| [OIII] | 4363.21 | 2.78 ± 0.35 | 2.84 ± 0.37 | 1.84 ± 0.24 | 1.88 ± 0.25 | 1.64 ± 0.24 | 1.72 ± 0.26 | 1.26 ± 0.15 | 1.31 ± 0.16 |
| Hβ | 4861.33 | 100.00 ± 0.58 | 100.00 ± 2.15 | 100.00 ± 0.56 | 100.00 ± 2.18 | 100.00 ± 0.41 | 100.00 ± 1.65 | 100.00 ± 0.42 | 100.00 ± 1.71 |
| [OIII] | 4958.92 | 141.05 ± 1.00 | 140.47 ± 4.19 | 99.58 ± 0.71 | 99.24 ± 3.02 | 81.17 ± 0.46 | 80.59 ± 1.86 | 91.76 ± 0.51 | 91.13 ± 2.18 |
| [OIII] | 5006.84 | 418.30 ± 3.02 | 415.64 ± 12.33 | 294.67 ± 2.24 | 293.11 ± 8.89 | 242.41 ± 1.53 | 239.72 ± 5.53 | 273.50 ± 1.71 | 270.57 ± 6.48 |
| HeI | 5875.67 | 12.28 ± 0.24 | 11.81 ± 0.39 | 12.95 ± 0.18 | 12.54 ± 0.38 | 13.18 ± 0.14 | 12.31 ± 0.28 | 12.66 ± 0.14 | 11.85 ± 0.29 |
| [NII] | 6548.03 | 4.88 ± 0.47 | 4.60 ± 0.46 | 7.09 ± 0.58 | 6.75 ± 0.58 | 9.66 ± 0.49 | 8.72 ± 0.47 | 7.80 ± 0.54 | 7.07 ± 0.51 |
| Hα | 6562.8 | 303.46 ± 1.97 | 286.00 ± 7.42 | 300.48 ± 1.99 | 286.00 ± 7.57 | 317.31 ± 1.59 | 286.00 ± 5.72 | 316.16 ± 1.65 | 286.00 ± 5.96 |
| [NII] | 6583.41 | 15.30 ± 0.48 | 14.41 ± 0.58 | 22.40 ± 0.60 | 21.31 ± 0.78 | 30.07 ± 0.51 | 27.08 ± 0.69 | 21.79 ± 0.55 | 21.79 ± 0.66 |
| [SII] | 6716.47 | 13.96 ± 0.16 | 13.10 ± 0.36 | 19.92 ± 0.21 | 18.90 ± 0.52 | 28.14 ± 0.22 | 25.18 ± 0.52 | 20.99 ± 0.18 | 18.86 ± 0.41 |
| [SII] | 6730.85 | 10.37 ± 0.15 | 9.73 ± 0.28 | 14.73 ± 0.19 | 13.96 ± 0.40 | 20.39 ± 0.20 | 18.24 ± 0.39 | 15.24 ± 0.17 | 13.68 ± 0.31 |
| [OII] | 7318.92 | 1.80 ± 0.06 | 1.67 ± 0.07 | 2.09 ± 0.07 | 1.95 ± 0.08 | 2.28 ± 0.11 | 1.99 ± 0.10 | 2.09 ± 0.11 | 1.84 ± 0.10 |
| [OII] | 7329.66 | 1.34 ± 0.05 | 1.24 ± 0.06 | 1.70 ± 0.06 | 1.59 ± 0.07 | 1.87 ± 0.10 | 1.63 ± 0.09 | 1.64 ± 0.09 | 1.44 ± 0.08 |
| E(B-V) | | 0.057 ± 0.006 | | 0.048 ± 0.006 | | 0.099 ± 0.005 | | 0.096 ± 0.005 | |
| F(Hβ) | | 28.46 ± 0.17 | 34.37 ± 0.74 | 25.20 ± 0.14 | 29.50 ± 0.64 | 19.31 ± 0.08 | 26.90 ± 0.44 | 87.11 ± 0.36 | 119.92 ± 2.06 |

Notes: F(Hβ) in units of × $10^{-15}$ erg cm$^{-2}$ s$^{-1}$

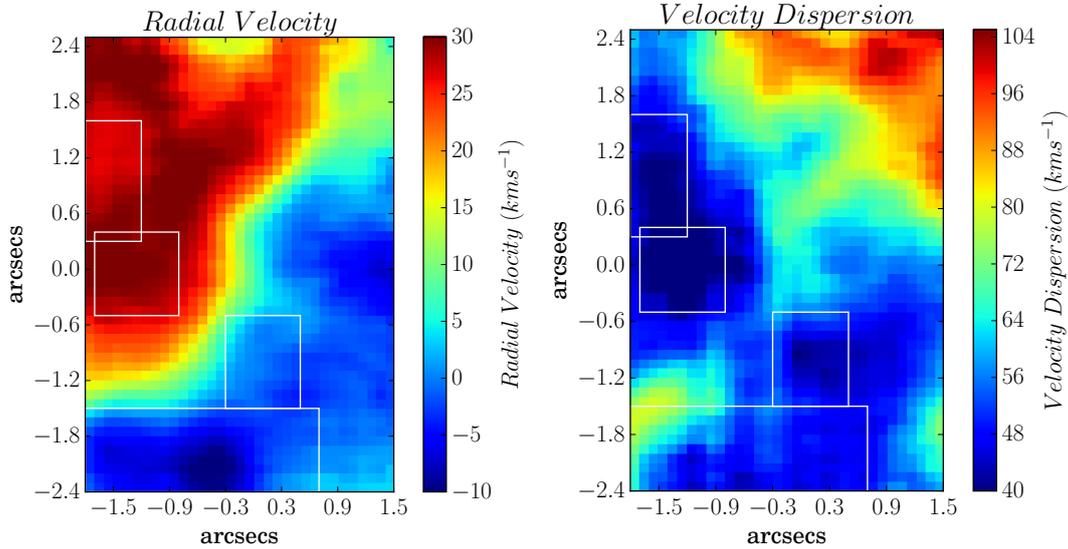

**Figure 7.** Radial velocity (left) and FWHM (right) maps of the ionised gas obtained from the Hα emission line. Radial velocity is corrected for systemic (= 1069 km s$^{-1}$) and barycentric velocities (= −13.74 km s$^{-1}$). FWHM is corrected for instrumental broadening (= 1.7Å). The four white rectangular boxes denote the location of the four H II regions.

while it is blueshifted in regions 3 and 4. Our velocity map shows a isovelocity S-shaped contour, which is in agreement with the [O III] velocity map of the entire galaxy presented by Cairós et al. (2012). Note here that their one spatial element covers our entire FOV.

The FWHM map (Figure 7, right panel) shows a variation of 40–104 km s$^{-1}$ across the FOV [3]. All the H II regions have a relatively lower velocity dispersion compared to the rest of the FOV.

### 3.3 Emission line ratio diagnostics

Figure 8 shows the classical emission line ratio diagnostic diagrams ([O III]$\lambda$5007/Hβ versus [S II]$\lambda\lambda$6717,6731/Hα (right panel) and [O III]$\lambda$5007/Hβ versus [N II]$\lambda$6584/Hα (left panel)), commonly known as BPT diagrams (Baldwin et al. 1981) which present a powerful tool to identify the ionisation mechanisms at play in the ionised gas. On both diagnostic diagrams, the solid black line represents the maximum starburst line, known as the "Kewley line" (Kewley et al. 2001) showing classification based on excitation mechanisms. The emission line ratios lying below and to the left of the Kewley line can be explained by the photoionisation by massive stars, while some other source of ionisation (e.g. AGNs or mechanical shocks) is required to explain the emission line ratios lying above the Kewley line. On the [N II]$\lambda$6584/Hα diagnostic diagram (Figure 8, right panel), the dashed black curve indicates the empirical line derived by Kauffmann et al. (2003) based on the SDSS spectra of 55 757 galaxies. The zone enclosed between this empirical curve and the theoretical "Kewley line" is referred to as the composite zone. The emission line ratios corresponding to the four H II regions (see Table 4) derived from their integrated spectra are shown as blue circle (region 1), green pentagon (region 2), red square (region 3) and or-ange diamond (region 4). The spatially-resolved (spaxel-by-spaxel) emission line ratios in the FOV for different regions are shown using the same colour and markers but smaller sizes. In addition, we also show the line ratios in the spaxels which are not covered by any of the four star-forming regions by magenta colored markers. We find that both the integrated and spatially-resolved data lie below and to the left of the Kewley line. Evidently no data lies in the composite region, hence confirming photoionisation by massive stars as the dominant ionisation mechanism in the target region of NGC 4670. Note here that the composite zone may also be assigned to pure H II regions with very high N/O (Pérez-Montero & Contini 2009). Although we see some some local N pollution in and around some H II regions in our sample (Figure 11), this is not enough to make our regions lie in the composite zone. Previous IFS studies of resolved H II regions in BCDs have shown that low-excitation spaxels can lie in the AGN region (see e.g. Pérez-Montero et al. (2011) and Kumari et al. (2017) for BCDs HS0128+2832, HS 0837+4717 and NGC 4449 respectively). However this is not the case for the spatially-resolved data in the BCD under study, NGC 4670.

---

[3] One may also estimate the thermal contribution to the velocity dispersion (Amorín et al. 2012b). From the mean $T_e$ values of the four H II regions (Table 4), we estimated this contribution as, $\sqrt{\frac{kT_e}{m_H}} \sim 9$ km s$^{-1}$. This correction leads the velocity dispersion to decrease by only ∼ 1 km s$^{-1}$.

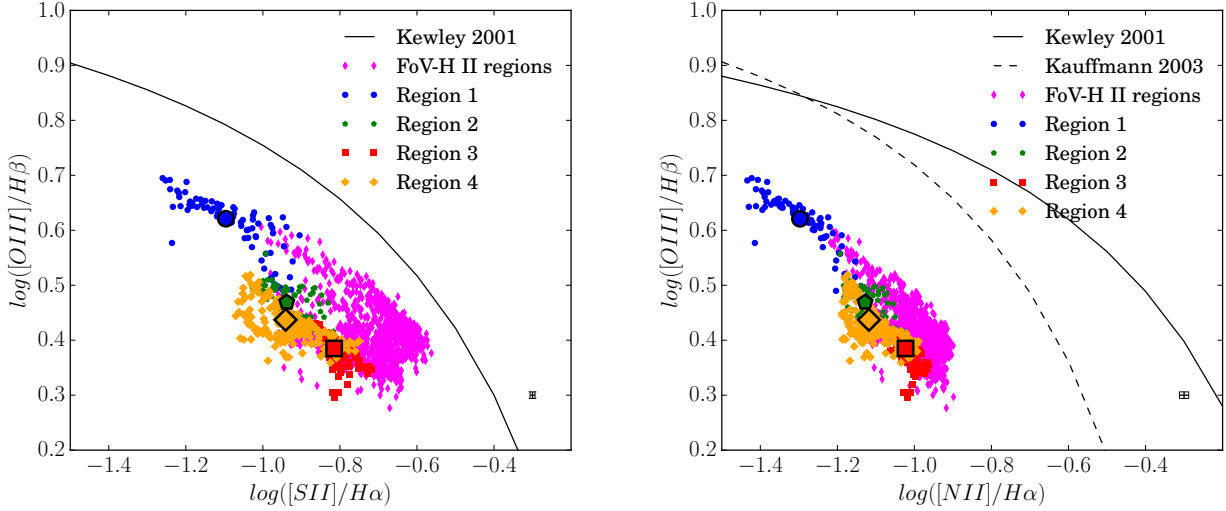

**Figure 8.** Emission line ratio diagnostic diagrams: [O III]/Hβ versus [S II]/Hα (left), and [O III]/Hβ versus [N II]/Hα (right). Black solid curve and dashed curve represent the theoretical maximum starburst line from Kewley et al. (2001) and Kauffmann et al. (2003) respectively, showing a classification based on excitation mechanisms. The line ratios of the four H II regions are colour-coded as follows: region 1: blue circle, region 2: green pentagon, region 3: red square, region 4: orange diamond. Smaller markers denote the spatially-resolved (spaxel-by-spaxel) line-ratios and the bigger markers denote the line-ratios obtained from the integrated spectrum of the corresponding regions. Magenta coloured markers denote the spatially-resolved line-ratios of the regions of FOV excluding the four H II regions. The size of errorbars varies for line ratios and the median error bars are shown in the right corner of each panel. The error bars on the line ratios obtained from the integrated spectra of the four H II regions are smaller than the markers used here.

**Table 4.** Integrated properties of the four H II regions of NGC 4670 in the GMOS-FOV

| | Ionisation Conditions | | | |
| --- | --- | --- | --- | --- |
| | Region 1 | Region 2 | Region 3 | Region 4 |
| log ([OIII]λ5007/Hβ) | 0.621 ± 0.003 | 0.469 ± 0.003 | 0.385 ± 0.003 | 0.437 ± 0.003 |
| log ([NII]λ6583/Hα) | -1.297 ± 0.013 | -1.128 ± 0.011 | -1.023 ± 0.007 | -1.118 ± 0.01 |
| log ([SII]λλ6717,6731/Hα) | -1.096 ± 0.004 | -0.938 ± 0.003 | -0.815 ± 0.003 | -0.941 ± 0.003 |
| | Abundance Analysis | | | |
| | Region 1 | Region 2 | Region 3 | Region 4 |
| $T_e$(O III) (K) | 10100 ± 400 | 9900 ± 400 | 10200 ± 500 | 9200 ± 300 |
| $T_e$(O II) (K) | 10500 ± 400 | 11000 ± 400 | 9700 ± 300 | 10000 ± 200 |
| $N_e$ (cm$^{-3}$) | 80 ± 30 | 70 ± 30 | <50 | <50 |
| 12 + log($O^+/H^+$) | 7.58 ± 0.06 | 7.55 ± 0.07 | 7.89 ± 0.06 | 7.77 ± 0.05 |
| 12 + log($O^{++}/H^+$) | 8.17 ± 0.06 | 8.05 ± 0.06 | 7.91 ± 0.07 | 8.12 ± 0.05 |
| 12 + log (O/H) | 8.27 ± 0.05 | 8.17 ± 0.05 | 8.21 ± 0.05 | 8.28 ± 0.04 |
| log(N/O) | -1.16 ± 0.07 | -1.02 ± 0.08 | -1.11 ± 0.07 | -1.11 ± 0.06 |
| $y^+$ (He I 5876) | 0.088±0.003 | 0.093±0.003 | 0.092±0.002 | 0.086±0.002 |
| | Stellar Properties | | | |
| | Region 1 | Region 2 | Region 3 | Region 4 |
| EW(Hα) (Å) | ~97 | ~152 | ~334 | ~265 |
| Age (sub-$Z_\odot$) (Myr) | ~24 | ~22 | ~16 | ~17 |
| SFR ($Z_\odot$)(×$10^{-3} M_\odot yr^{-1}$) | 20.3 ± 0.3 | 17.5 ± 0.3 | 15.9 ± 0.2 | 70.9 ± 0.8 |
| SFR (sub-$Z_\odot$)(×$10^{-3} M_\odot yr^{-1}$) | 13.4 ± 0.2 | 10.8 ± 0.2 | 10.1 ± 0.1 | 47.1 ± 0.6 |

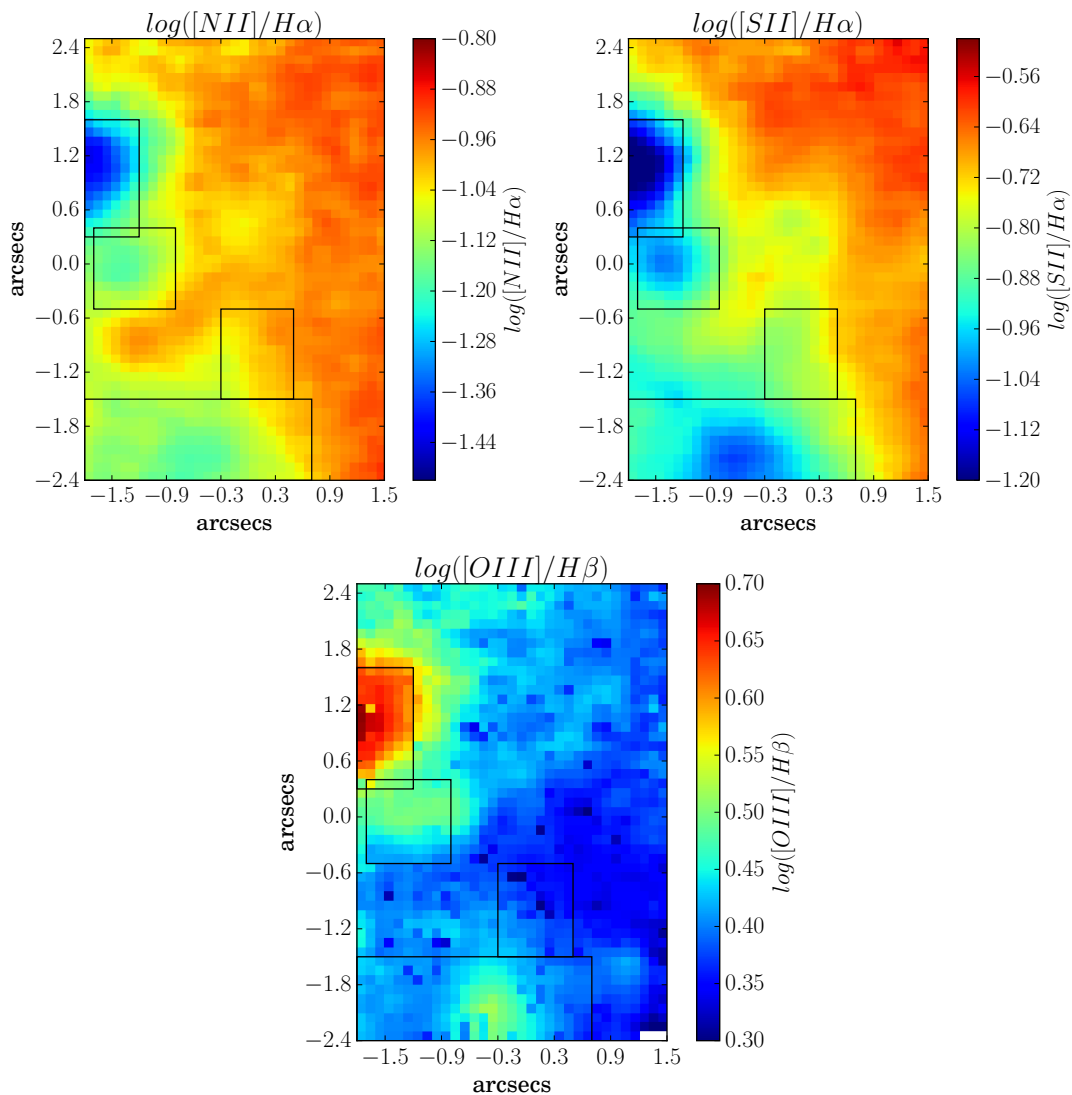

**Figure 9.** Emission line ratio maps of $[NII]\lambda6583/H\alpha$, $[SII]\lambda\lambda6717,6731/H\alpha$ and $[OIII]\lambda5007/H\beta$. The four black rectangular boxes denote the location of the four H II regions. White spaxels correspond to those spaxels with emission line fluxes having S/N < 3.

We can study the spatial structure of the ionisation through the relevant line ratio maps shown in Figure 9. The H II regions, particularly Regions 1, 2 and 4 show low values of [N II]$\lambda$6584/H$\alpha$ (upper-left panel) and [S II]$\lambda\lambda$6717,6731/H$\alpha$ (upper-right panel), and high values of [O II]$\lambda$5007/H$\beta$ (lower panel), indicating that the corresponding regions have relatively high excitation. This is likely due to the presence of a harder ionising field from hot massive stars in the respective H II regions. These regions are also brighter in the B-band continuum (Figure 5, upper-left panel), supporting our inference of the presence of massive stars.

### 3.4 Chemical Abundances

#### 3.4.1 Integrated spectra chemical abundances

We estimate chemical abundances of the four H II regions in the target region of NGC 4670 from their integrated spectra (Figure 3) by the direct method (i.e. using electron temperatures and collisionally excited lines). All emission line fluxes are reddenning-corrected for chemical abundance determination. Table 4 summarises the chemical properties of the four H II regions.

**Electron Temperature and Density:** The first step in estimating chemical abundance by the direct method is the determination of electron temperature($T_e$) and density ($N_e$). We derive $T_e$ ([O III]) from the dereddenned [O III] line ratio, [O III] ($\lambda$5007 +$\lambda$4959)/[O

iii]$\lambda$4363 and the expression from Pérez-Montero (2017), which is obtained assuming a five-level atom, using collision strengths from Aggarwal & Keenan (1999), and is valid in the range 7000-25000 K. The estimated $T_e$ ([O iii]) for the four H ii regions vary from ~9200–10200 K (Table 4).

Using the derived $T_e$ ([O iii]) value for each star-forming region and the corresponding [S ii] doublet ratio $\lambda$6717/$\lambda$6731 from the integrated spectra, we compute $N_e$ ([S ii]) of the four H ii regions (Table 4). We find $N_e$ to be low in all the four H ii regions, with values < 50 cm$^{-3}$ in regions 3 and 4. Such low densities and derived $T_e$ ([O iii]) are common within H ii regions (Osterbrock & Ferland 2006).

To derive the temperature of the low-ionisation zone $T_e$ ([O ii]), we employ the expression given in Pérez-Montero (2017)[4], which uses the ratio of oxygen doublets, ( [O ii] ($\lambda$3726 + $\lambda$3729)/[O ii]( $\lambda$7319 + $\lambda$7330)), in combination with the electron density $N_e$ (derived above). The expression is valid in the range 8000–25000K, where the collision coefficients are taken from Pradhan et al. (2006) and Tayal (2007). The estimated $T_e$ ([O ii]) of the four H ii regions vary from ~9700-11000 K (Table 4).

**Oxygen abundance:** Oxygen is used as a proxy for total metallicity because it is the most prominent heavy element observed in the optical spectrum in the form of $O^0$, $O^+$, $O^{2+}$, $O^{3+}$. We employ the formulations of Pérez-Montero (2017) to calculate $O^+/H^+$ and $O^{2+}/H^+$ using $T_e$ ([O ii]) and $T_e$ ([O iii ]) respectively, i.e. the electron temperatures of the ionisation zone dominated by the corresponding ions. The $O^+/H^+$ and $O^{2+}/H^+$ are combined to calculate the elemental O/H for all four H ii regions given in Table 4. The values of 12+log(O/H) vary between 8.17–8.28, with a mean = 8.23 and standard deviation = 0.04. These values fall in the transition region between the intermediate- and the high-metallicity regime which we discuss further in Section 4.

Previous studies reported the metallicity of NGC 4670 using only indirect methods. For example, Mas-Hesse & Kunth (1999) estimates 12+log(O/H) = 8.4 from a 11″slit spectrum , while Cairós et al. (2012), calculated metallicity using another strong-line calibration, the P-method (Pilyugin & Thuan 2005) and reported 12 + log(O)/H = 8.29 for the entire galaxy (~ 80″ × 80″) and 12+log(O/H) = 8.37 for the nuclear region (~ 40″ × 40″). The mean metallicity of 12+log(O/H) = 8.23 calculated from the direct method in this work is lower than all those values, which is consistent with the systematic offsets existing between different metallicity diagnostics as noted by Kewley & Ellison (2008). Moreover the region under study in this work is much smaller (~ 3.5″ × 5″) than the previous studies.

**Nitrogen-to-oxygen ratio:** Assuming the low-ionisation zone temperature $T_e$ ([O ii]) and density $N_e$ ([S ii]) derived above, we use the emission line ratio of ([N ii]($\lambda$6584 + $\lambda$6584)/ H$\beta$ formulation from Pérez-Montero (2017) to derive log($N^+/H^+$) for all four H ii regions. Assuming that $N^+/O^+$ = N/O, we use log($N^+/H^+$) and log($O^+/H^+$) to derive log(N/O). These values agree with those derived directly from ([N ii]$\lambda$6584/ [O ii] ($\lambda$3726 + $\lambda$3729)) using the log(N/O) formula from Pérez-Montero (2017)[4]. The log(N/O) values of the four H ii regions vary between −1.02– −1.16, with a mean of −1.10 and a standard deviation of 0.05.

**Helium abundance:** We use the reddenning-corrected He i 5876 line in combination with $T_e$ ([O iii]) and $N_e$ ([S ii]) derived above, to calculate the helium abundances ($y^+$) of the four H ii regions. Other helium lines are not used as they are either weak (S/N

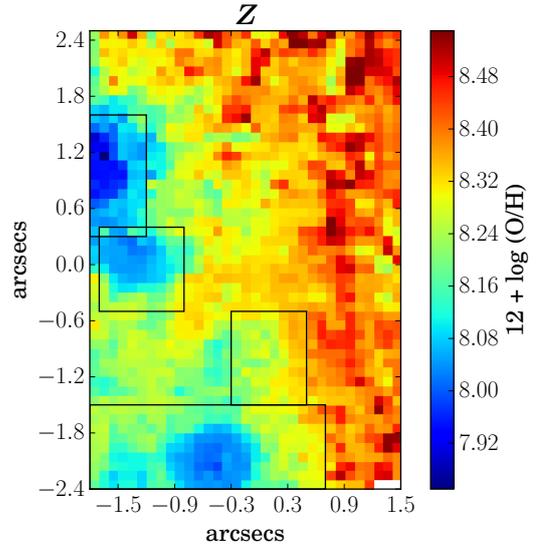

**Figure 10.** Metallicity map obtained using the HII-CHI-mistry code. The four black rectangular boxes denote the location of the four H ii regions. White spaxels correspond to the spaxels in which emission line fluxes had S/N < 3.

< 3) or affected by absorption of underlying stellar populations in one or more H ii regions. The expression for theoretical emissivities are taken from Pérez-Montero (2017), and the optical depth function is assumed to be one. This assumption on optical depth along with the contamination of He lines due to underlying absorption feature would result in an uncertainty of only ~2% in the helium abundance (Hägele et al. 2008). The $y^+$ values for the four H ii regions vary between 0.086–0.093, with a mean of 0.090 and standard deviation of 0.003.

### 3.4.2 Spatially-resolved chemical abundances

To map chemical abundance using the robust $T_e$-method, we require the detection of a weak auroral line throughout the FOV. Unfortunately, the S/N ratio of [O iii]$\lambda$4363 in the individual spectrum across the FOV was too low (< 3) to use the direct $T_e$-method. Various indirect methods involving the use of strong emission lines have been devised to estimate chemical abundances to overcome this problem. Some of the popular indirect methods involve the use of the N2 parameter (Denicoló et al. 2002), O3N2 parameter (Pettini & Pagel 2004), $R_{23}$ (Pagel et al. 1979) N2S2H$\alpha$ (Dopita et al. 2016) or a combination of them (e.g. Maiolino et al. 2008; Curti et al. 2017). We used three of these strong line methods in our analysis of another BCD, NGC 4449 (Kumari et al. 2017) and found the well-known problem of the offsets between the metallicities from the direct and indirect method Kewley & Ellison (2008). In this work, instead, we make use of the publicly-available python-based code HII-CHI-mistry (v3.0)[5]. This code takes the dereddenned fluxes, follows a $\chi^2$-based methodology on a grid of photoionisation models, including CLOUDY and POPSTAR, and outputs chemical-abundances (O/H, N/O) (Pérez-Montero 2014). The errors on the output abundances are calculated by the code using a Monte Carlo iteration from the reported errors of the measured lines. By using this code, we aim to remove the dependency of oxygen abundance on the nitrogen-to-oxygen ratio, which is an inherent assumption in some of the strong line calibrators. This becomes useful when we study the spatially-resolved (O/H) versus (N/O) later in Section 4.

---

[4] See Appendix A for updated formulae.

[5] http://www.iaa.es/ epm/HII-CHI-mistry.html

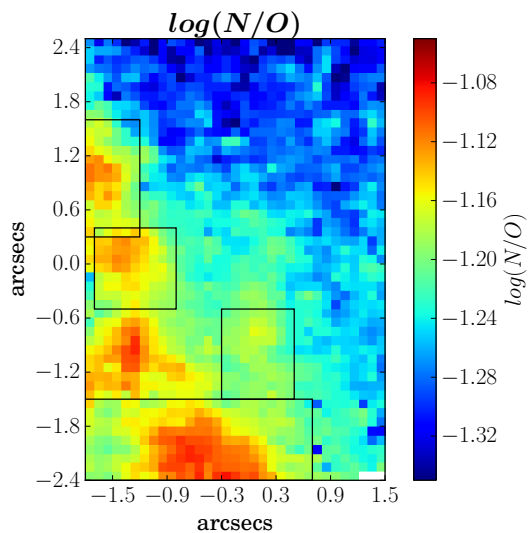

**Figure 11.** log(N/O) map obtained using the HII-CHI-mistry code. The four black rectangular boxes denote the location of the four H II regions. White spaxels correspond to the spaxels in which emission line fluxes had S/N < 3.

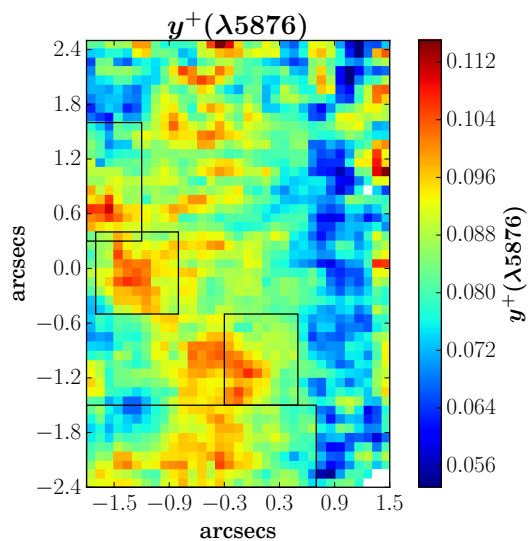

**Figure 12.** Helium abundance map obtained from He I $\lambda$5876 flux map. The four black rectangular boxes denote the location of the four H II regions. White spaxels correspond to the spaxels in which emission line fluxes had S/N < 3.

**Oxygen abundance:** Figure 10 shows the metallicity map of the FOV obtained from the HII-CHI-mistry code. The value of 12 + log(O/H) varies between ~ 7.80–8.56 with a mean value of = 8.29 and standard deviation = 0.13. We find that region 1 is metal-poor compared to the other three H II regions. However, metallicity estimates obtained from the integrated spectra of the four H II regions show that regions 1 and 4 have comparable metallicities. This is likely due to an aperture effect, i.e. when we integrate over the large regions, the temperature and metallicity gets dominated by the brightest regions and becomes luminosity-weighted. Thus even though region 4 hosts a range of metallicities, it is dominated by the large low-metallicity blob which has similar metallicity to that of region 1 as inferred from the integrated spectra. The mean value of spatially-resolved log(O/H) agrees with the mean value calculated from the integrated spectra of the four H II regions within ±1 standard deviation.

**Nitrogen-to-oxygen ratio:** Figure 11 shows the log(N/O) map of the FOV from the HII-CHI-mistry code. The value of log(N/O) varies between −1.4 − −1.08, with a mean value of = -1.23 and standard deviation = 0.06. We find that there is an increase in log(N/O) values in the region surrounded by the three H II regions 2,3 and 4. This is probably a region which formed stars in the past but is quiescent now, resulting in a relatively chemical-enriched ionised gas. The log(N/O) map also shows that region 3 has a relatively lower log(N/O) value compared to other H II regions, however it is the integrated spectrum of region 1 which shows the highest log(N/O) (Table 4). This is likely due to aperture effects as explained above. The mean value of spatially-resolved log(N/O) agrees with the mean value calculated from the integrated spectra of the four H II regions within ±2 standard deviations.

The maps of log(O/H) and log(N/O) show signatures of chemical inhomogeneity though the significance is low because of high error bars. We explore the homogeneity of chemical abundances in an upcoming paper (Kumari et al 2018 (in prep)).

**Helium abundance:** For mapping helium abundance ($y^+$), we require spatially-resolved electron temperature and density measurements. As mentioned earlier, such maps could not be made because of the low S/N of the weak auroral line [O III] $\lambda$4363. So we create the $T_e$ ([O III]) map by using the HII-CHI-mistry derived metallicity map (Figure 10), and the relation between log(O/H) and $T_e$ ([O III]) from Pérez-Montero (2017) (proposed initially in Amorín et al. (2015)). We use the $T_e$ ([O III]) map in conjunction with spatially-resolved [S II] doublets to map $N_e$ across the FOV [6]. Among all the He lines detected in our data (He I 4471, 5876, 6678 and 7065Å , He II 4686), the flux map corresponding to He I $\lambda$5876 has the highest $S/N$ across the entire FOV. Hence, we use the dereddenned flux map of He I $\lambda$5876 along with the $T_e$ ([O III]) and $N_e$ maps to create the helium abundance map (Figure 12). The theoretical emissivities and the optical depth function are same as described in Section 3.4.1. Comparing the helium abundance map (Figure 12) with the B-band continuum (Figure 5), we find that there is a relative increase of $y^+$ in the regions surrounding the continuum. Given that the continuum indicates the region with a relatively older population, it is likely that the winds emanating from the outer atmosphere of these stars or cluster of stars have resulted in an increase in $y^+$ in the surrounding regions.

### 3.5 Stellar Properties

#### 3.5.1 Age of stellar population

The integrated spectra of the four H II regions (Figure 3) show Balmer emission lines indicating the presence of young, hot and massive O and B stars, and Balmer absorption lines (H$\beta$, H$\gamma$ etc.) indicating the presence of early-type A stars. For age-dating the current ionising population, we first map the equivalent width (EW) of Hα recombination line as shown in Figure 13. The EW values show a variation of 48–776 Å across the FOV. Such high values of EW(Hα) have been found in H II regions in the local BCDs (e.g. I Zw 18, Mrk 475, Mrk 1236, NGC 2363 by Buckalew et al. 2005). Considering higher redshifts, Shim et al. (2011) found EW(Hα) = 140–1700 Å for a galaxy sample in the redshift range of 3.8 < z < 5.0, although with higher star-formation rates (∼20–500 M$_\odot$yr$^{-1}$). High EW(Hα) (697–1550Å) are also found in the extreme green peas (GPs) (Jaskot & Oey 2013). Thus NGC 4670 may be similar to the GPs and high redshift galaxies, with respect to EW(Hα).

---

[6] The maps of $T_e$ ([O III]) and $N_e$ derived in this section are presented in Appendix B

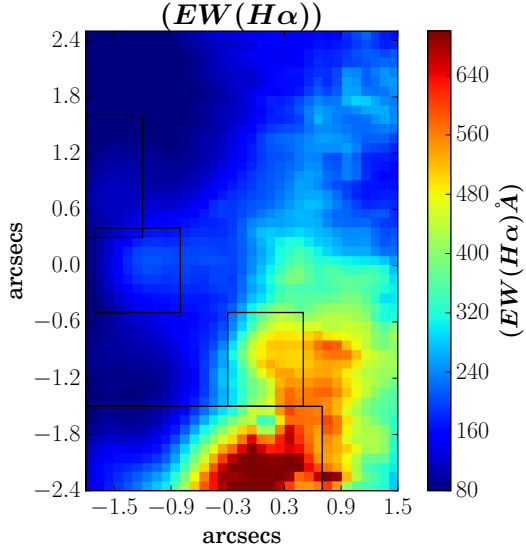

**Figure 13.** Map of equivalent width of Hα. The four black rectangular boxes denote the location of the four H II regions.

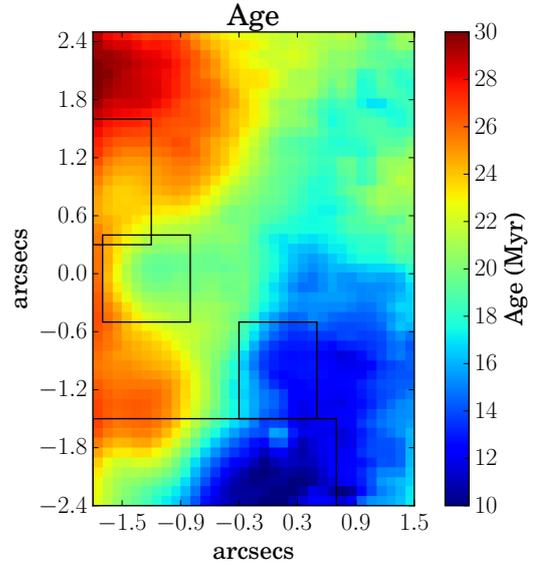

**Figure 14.** Age map in Myr calculated from Starburst99 models at the constant metallicity of Z = 0.35 Z$_\odot$ (mean of metallicities of the four H II regions tabulated in Table 4).

To find the age of the stellar population, we next estimate the EW(Hα) from the evolutionary synthesis models of Starburst99 (Leitherer et al. 1999) at a constant mean metallicity (∼ 0.35 Z$_\odot$) of the four H II regions (calculated in Section 3.4.1). For generating models corresponding to the properties typical of H II regions, we adopt Geneva tracks with standard mass-loss rates assuming instantaneous star-formation. The choice of evolutionary tracks results in a relatively small change in the estimated age, with Padova tracks predicting higher ages by upto 20% (James et al. 2010). We assume a Salpeter type initial mass function (IMF) (Salpeter 1955) and the total stellar mass extent between the upper and lower cut-offs fixed to default value of 10$^6$ M$_\odot$. The expanding stellar atmosphere is taken into account by using the recommended and realistic models of Pauldrach/Hiller. We have also taken into account the stellar rotation in our models though that leads to an insignificant increase of 3% in age (Kumari et al. 2017). We compare the modelled EW with the observed EW (Figure 13) and, hence map the corresponding age shown in Figure 14 which shows a variation of 10–30 Myr across the FOV. The map shows that the stellar population of approximately same age is present in each star-forming region, except region 4 where stellar population of different ages are present, with the youngest at the peak of region 4. The older-age stellar population also seems to align with the two peaks of B-band continuum map (Figure 5, upper-left panel). An interesting observation is the similarity of the age map and the velocity map (Figure 7, left panel), which probably indicates that the ionised gas near the same-age stellar population have similar velocities.

We also measure EW(Hα) from the integrated spectra of the four H II regions and calculate the age of the stellar population residing in those regions at sub-solar metallicities (found for each region). The results are tabulated in Table 4. The age of these regions are in agreement with the spatially-resolved age shown in Figure 14.

The age range of 10–30 Myr derived here, is higher than the age of ∼4 Myr estimated by the Mas-Hesse & Kunth (1999) from their evolutionary population synthesis models. The difference is mainly because the models of Mas-Hesse & Kunth (1999) are weakly sensitive to previous star-formation episodes but rather to the ones younger than 10 Myr. Their calculation had been for a period of only 20 Myr since they were mainly interested in the evolution of massive stars. In contrast, the upper age limit on our models is 100 Myr. In spite of all evident explanations of the difference in results, we caution the readers to interpret our results in light of the various systematic uncertainties involved in our modelling, e.g. those related to the internal dust attenuation, determination of continuum level which is the combination of the nebular and stellar continuum resulting from different age stellar populations existent in the same region (see e.g. Cantin et al. 2010; Pérez-Montero & Díaz 2007).

### 3.5.2 Wolf-Rayet Stars

Figure 15 shows the emission map of the blue WR feature showing the distribution of WR stars in the FOV. The map is created by integrating over the full emission feature from 4600-4700 Å after subtracting the underlying continuum. The blue bump of the WR feature is mainly composed of N V, N III, C III/IV blends and the He II λ4686 line, and is generally contaminated by nebular line [Fe III] λ4658 line. However this contaminating line is not strong enough (S/N < 3) at the spatially-resolved scale in our data[7]. Hence we do not remove its contribution while mapping the WR feature shown in Figure 15. In this map, all white spaxels correspond to the spaxels in which the combined WR blend or feature has S/N < 3, and the black dashed rectangular boxes denote the location of the four H II regions. We find that the WR feature becomes prominent in the region to the south-east of the peak of region 4, and to the south-west of region 2. We mark the corresponding region (WR region) of peak WR emission by the red rectangular box in Figure 15. This region do not contain the peak of any of the four H II regions but lies close to regions 2 and 4. This may indicate the propagation of star-formation from the current WR region to the current H II regions and has been observed before in a dwarf galaxy Mrk 178 (Kehrig et al. 2013). The region of prominent WR emission also shows an increase in log(N/O) (Figure 11). This observation is expected because WR stars are the evolved phases of massive O stars which lose N to the interstellar medium via stellar winds. These winds may also explain the slight increase in the $y^+$ map (Figure 12) surrounding the WR region. The age of WR stars are expected to be in the range ∼3–8 Myr (Meynet 1995). However our age map (Figure 14) reveals a slightly older (∼10–30 Myr) ionising stellar population in this region. This simply indicates the existence of an older stellar population along with the younger WR stars.

---

[7] Taking the ratio of summed fluxes of WR map with S/N>3 and summed fluxes of [Fe III] λ4658 with S/N>1, we find an upper-limit of 6% as contribution of [Fe III] λ4658 in the WR map.

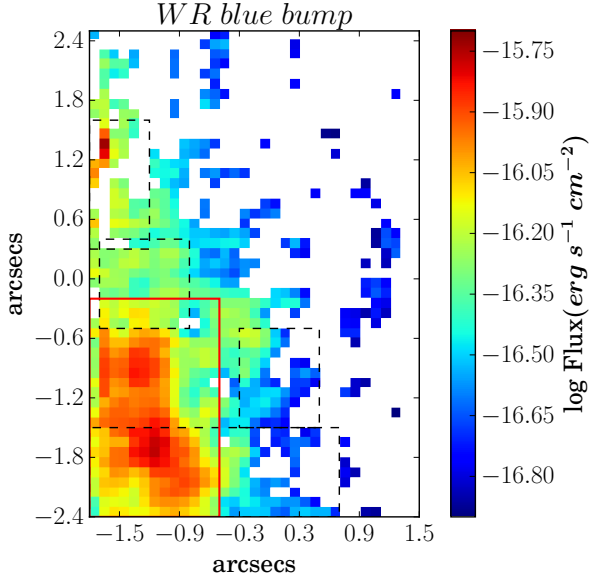

**Figure 15.** Emission map of the blue WR feature (created by integrating over the full emission feature from 4600–4700Å), showing the distribution of WR stars in the FOV. The red rectangular box shows the peak of the WR distribution which we use for subsequent analysis. The four black dashed rectangular boxes denote the location of the four H II regions. White spaxels correspond to the spaxels in which WR emission fluxes had S/N < 3.

Figure 16 presents the WR "blue bump" feature in integrated spectra of all H II regions along with the WR region, which shows the peak in the WR distribution (red box in Figure 15). In all these spectra, we find a gap between the broad 4640 feature and the [Fe III] $\lambda$4658.05 emission line. This rare feature was first identified strongly in NGC 3049 (Schaerer et al. 1999) and explained as a real feature by Schmutz & Vacca (1999). Basically, the gap arises because the 4640 feature is not one broad feature but a combination of at least three emission components (N v $\lambda\lambda$4604, 4620, N III $\lambda\lambda$4634, 4641, C III $\lambda\lambda$4647, 4650). Except region 4, none of the star-forming regions show a prominent WR "blue bump" feature. Inspecting the spectra of region 4, we find the observed blue bump is mainly due to the common region which it shares with the WR

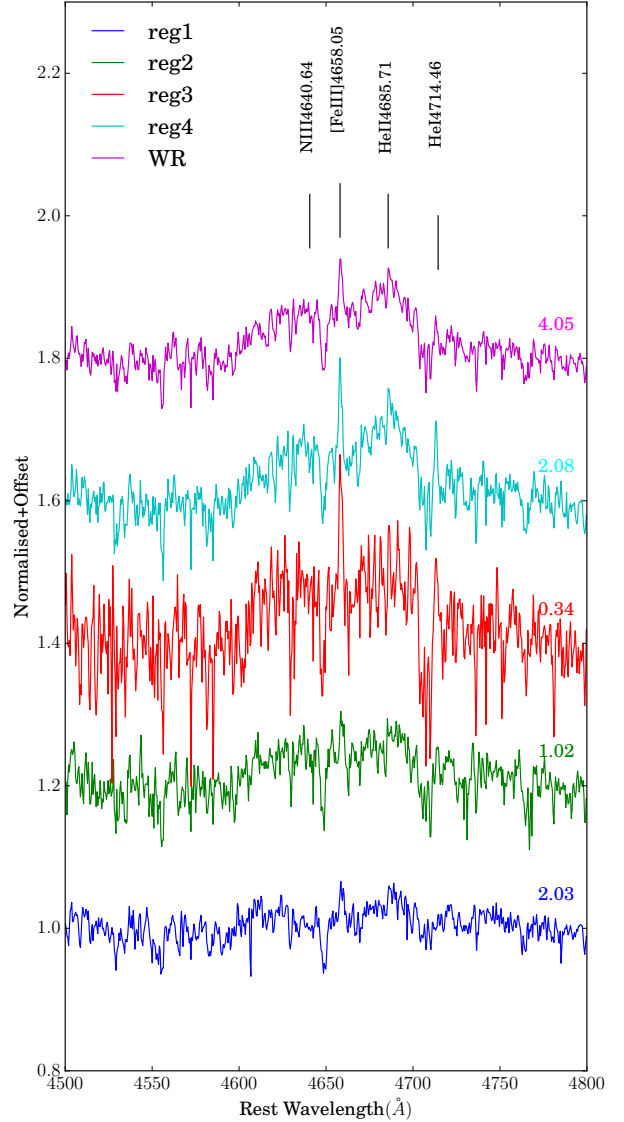

**Figure 16.** WR 'blue bump' features (N v $\lambda$4620, N III $\lambda$4640 and He II $\lambda$4685) in the integrated spectra of all H II regions along with the region showing the peak in WR distribution (red box in Figure 15). The integrated spectra are normalised by the continuum in the respective region and then offset by 0.2 with respect to each other for better visibility. On the right hand side above each spectrum, we show the average level of continuum (in units of $10^{-15}$ erg s$^{-1}$ cm$^{-2}$ Å$^{-1}$) in the given wavelength range corresponding to each region in the same colour. The blue bump indicates the presence of late-type WN and WC stars. All these spectra also show a gap between the broad $\lambda$4640 feature and the nebular iron line [Fe III] $\lambda$4658.05Å.

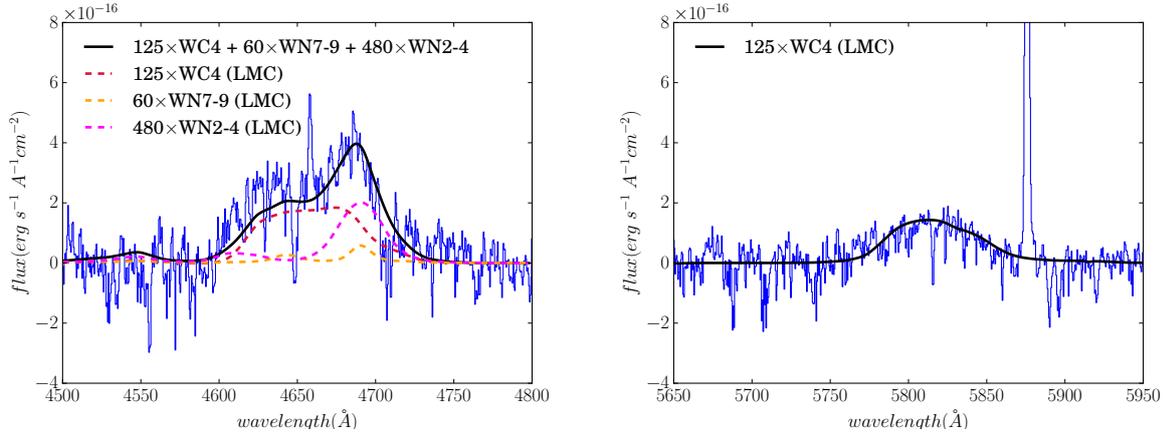

**Figure 17.** Left: Continuum-subtracted WR 'blue bump' feature summed over the peak of WR distribution (red box in Figure 15). Overplotted is the combined fit (solid black curve) for 125 WC4 stars (red dashed curve), 60 WN7-9 stars (orange dashed curve) and 480 WN2-4 stars (magenta dashed curve), using the LMC templates of Crowther & Hadfield (2006). Right: Continuum-subtracted WR 'red bump' feature summed over the peak of WR distribution (red box in Figure 15). Overplotted (solid black curve) is the scaled LMC template of WC4 star from Crowther & Hadfield (2006), denoting the presence of 125 WC4 stars.

region. Hence we concentrate on the WR region for subsequent analysis of WR stars.

We estimate the number of WR stars in the region showing the peak of the WR distribution (red box in Figure 15), by fitting WR templates to the integrated spectrum of the corresponding region. Since the mean metallicity of the FOV (12 + log(O/H) = 8.23) is closer to that of LMC than the Small Magellanic Cloud (i.e. 12 + log(O/H) = 8.35 and 12 + log(O/H) = 8.03 respectively, (Russell & Dopita 1992)), we use the WR LMC templates (Crowther & Hadfield 2006). The specific LMC templates are selected on the basis of the relative strength of different components of the WR blue and red bumps revealed by our spectrum. For modelling the red bump, both WC4 and WO templates are available. We did a $\chi^2$-minimization using the two templates separately for the red-bump, and found that the WC4 template produced a better fit. Moreover, our spectrum has a clear C IV $\lambda$5808 feature indicating the presence of WC4 stars, but no O V $\lambda$5990, a signature of WO stars. Hence we selected WC4 template for the rest of the template fitting. For fitting the more complex blue bump, we selected the WN7-9 template over the WN5-6 template, because the relative strength of He II $\lambda$4686 and and the broad emission ~4640Å in our spectrum are similar to WN7-9 template. We also include the WN2-4 template for modelling the N V $\lambda$4604, 4620 feature revealed in our spectrum. Finally, we perform a $\chi^2$-minimization using the above selected templates and simultaneously fit all spectral regions of interest, i.e. the red bump and the three clear features of blue bump (N V $\lambda\lambda$4604, 4620, N III $\lambda\lambda$4634, 4641, He II $\lambda$4686). In $\chi^2$-minimization, we used the dispersion in the emission-free continuum region as the uncertainty on the input WR fluxes. We calculated the uncertainty on the number of WR stars obtained from a Monte-Carlo simulation using the dispersion in the emission free region. Hence we estimate ~125±4 carbon-type (WC4) stars, ~60±20 late-type (WN7-9) stars and ~480±50 early-type (WN2-4) stars. The corresponding fit is shown in Figure 17 for both blue (left panel) and red bump (right panel). Cairós et al. (2012) reports WR signatures in the central region of NGC 4670 which approximately overlaps with our FOV, though there is no estimate of the number of WR stars available for comparison with our study. We have found ~670±50 WR stars over a region of 212 pc × 116 pc. Our density is lower (~ a factor of 2) in comparison to another BCD Mrk 996 where ~ 3000 WR stars are reported in an area of $4.6 \times 10^4$ pc$^2$ (James et al. 2009). However, these densities are higher in comparison to a nearby WR galaxy Mrk 178, where 20 WR stars are found in a region of ~ 300 × 230 pc (Kehrig et al. 2013).

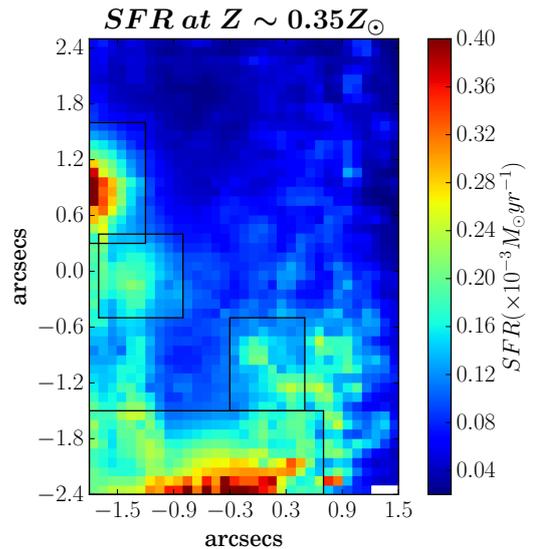

**Figure 18.** SFR Map of the FOV in units of × $10^{-3}$ M$_\odot$ yr$^{-1}$ at 0.35 Z$_\odot$. The four black rectangular boxes denote the location of the four H II regions.

### 3.5.3 Star Formation Rate

Assuming a constant metallicity throughout the FOV, we map the SFR at Z = 0.35 Z$_\odot$ (Figure 18) using the metallicity-dependent relation between dereddened luminosity of H$\alpha$, L(H$\alpha$) and SFR from Ly et al. (2016) derived from the Starburst99 models assuming a Padova stellar track and Chabrier IMF(Chabrier 2003). The SFR values vary between 0.02–0.49 × $10^{-3}$ M$_\odot$ yr$^{-1}$ at sub-solar metallicity. We also map the SFR at solar-metallicity using the L(H$\alpha$)–SFR relation given in Kennicutt (1998) for a Salpeter IMF.

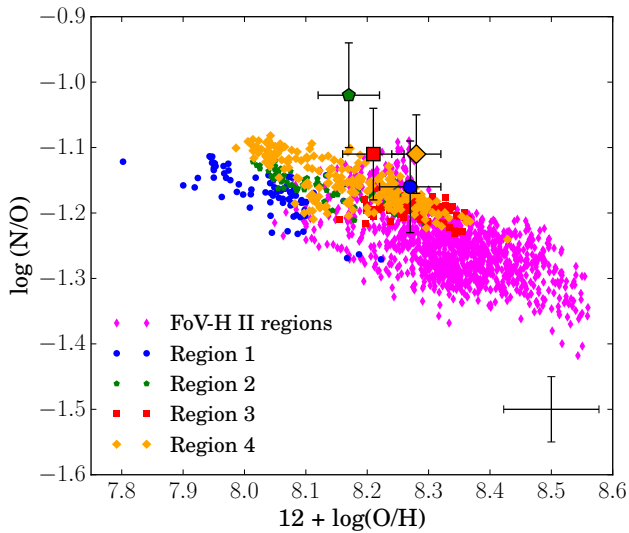

**Figure 19.** The relationship between log(N/O) versus 12 + log(O/H) for the four H II regions as well as the whole FOV (on spaxel-by-spaxel basis). The four regions are color-coded as follows: region 1: blue circle, region 2: green pentagon, region 3: red square, region 4: orange diamond. Smaller markers denote the spatially-resolved (spaxel-by-spaxel) relation where the two quantities (log(N/O) and 12 + log(O/H)) are calculated from the HII-CHI-mistry code whereas the bigger markers denote quantites calculated from the direct $T_e$-method for the four H II regions. Magenta coloured markers denote the spatially-resolved quantities of the regions of FOV excluding the four H II regions. The median error bars for the spatially-resolved quantities are shown in the right-hand corner.

We include a multiplicative factor of 0.63 in the SFR map at solar metallicity to account for the Chabrier IMF, which has a more realistic distribution at the lower end of stellar masses. We find that the SFR varies between $0.03-0.77 \times 10^{-3}$ $M_\odot$ yr$^{-1}$ at solar metallicity. The SFR map at solar metallicity shows a similar pattern as the one at sub-solar metallicity.

We also estimate the SFR for the four H II regions at solar and sub-solar metallicities for each region using the recipes mentioned above. The results are given in Table 4. For both integrated data of the four H II regions and spatially-resolved data, we find that the SFR at sub-solar metallicity is lower than those at solar metallicity. This is because the atmosphere of metal-poor O stars are more transparent resulting in an increased escape fraction for the ionising photons and hence lower SFR. Our results are in agreement with the SFRs of other BCDs, which span a range of $10^{-3}$ to $10^2$ $M_\odot$ yr$^{-1}$ (Hopkins et al. 2002).

## 4 DISCUSSION

### 4.1 N/O Conundrum revisited

The relationship between the nitrogen-to-oxygen ratio and the oxygen abundance in H II regions has been the subject of intense debate (e.g. see McCall et al. 1985; Thuan et al. 1995; Henry et al. 2000; Izotov et al. 2006; López-Sánchez & Esteban 2010; Vincenzo et al. 2016; Belfiore et al. 2017), and has been used to probe the origin of nitrogen in different metallicity regimes. Based on the analysis of 54 H II regions in BCDs, Izotov & Thuan (1999) proposed three metallicity regimes related to the origin of nitrogen. In the low-metallicity regime (12 + log(O/H) < 7.6), nitrogen is thought to have a primary origin and is produced by massive stars only, which set the level of log(N/O) at ∼ −1.6 with a very low scatter. However, since the mechanism of production of nitrogen in massive stars is not clearly understood, the behaviour of the relationship in the low-metallicity regime could also result due to effects related to stellar rotation (Meynet & Maeder 2005), low SFR (Henry et al. 2000), low number of WR stars (Izotov et al. 2006), or the production of nitrogen by low-and intermediate-mass stars (see e.g López-Sánchez & Esteban 2010). In the intermediate-metallicity regime (7.6 < 12 + log(O/H) < 8.3), the value of log(N/O) increases above −1.6 and the origin of nitrogen is interpreted to be primary but produced by intermediate-mass stars, which evolve and release their nucleosynthesis products into the interstellar medium. The relation shows a large scatter (±0.3 dex). Berg et al. (2012) reports a positive slope between the log(N/O) and 12 + log(O/H) for the metallicity values 12 + log(O/H) > 7.7, indicating a secondary origin of nitrogen even in the intermediate-metallicity regime. However, their analysis of H II regions in low-luminosity dwarf galaxies does show a large scatter. This scatter has been observed in several observational studies of systems with low and intermediate metallicities (e.g. van Zee & Haynes 2006; Pérez-Montero & Contini 2009; James et al. 2015, 2017), and has been attributed to factors such as, the time delay between the production of oxygen and secondary nitrogen (Garnett 1990; Vila Costas & Edmunds 1993; Pettini et al. 2008), variation in the star formation histories (Mollá et al. 2006; Mollá & Gavilán 2010), the global gas flows in galaxies (Köppen & Hensler 2005; Amorín et al. 2012b). At high metallicities (12 + log(O/H) > 8.3), nitrogen is thought to have a secondary origin, and hence a positive slope is expected between log(N/O) and 12 + log(O/H). Interestingly, Kobulnicky & Skillman (1996) report a negative trend in the relation for H II regions within the irregular starburst galaxy NGC 4214, with strong WR features.

In Figure 19, we present the relationship between log(N/O) versus 12 + log(O/H) for the four H II regions as well as the whole FOV (on spaxel-by-spaxel basis). We caution here that given the spatial scale of our data (∼9 pc), each spaxel may or may not host H II regions since the typical size of H II regions varies between a few to hundreds of pc (Kennicutt 1984). The oxygen abundance and the nitrogen-to-oxygen ratio for the four H II regions are calculated from the direct $T_e$-method. However, O/H and N/O for the spatially-resolved data are estimated from the HII-CHI-mistry code. For the four H II regions, we compare these quantities calculated from the direct $T_e$-method and the HII-CHI-mistry which included the use of [O III] $\lambda$ 4363 line. We find that they agree with each other within $1\sigma$, except for the metallicity of one H II region (Region 4) which agrees within $2\sigma$. As noted by Pérez-Montero (2014), there are dispersions of ∼0.2 dex in the HII-CHI-mistry model-based and direct $T_e$-method abundances. We expect no trend or a positive trend in these regimes as mentioned before. However, we find a hint of negative trend between N/O and O/H for the four H II regions, while the negative trend is more prominent for the spatially-resolved data (on spaxel-by-spaxel basis), covering 0.3 dex on y-axis and 0.6 dex on x-axis lying in the intermediate- and the high-metallicity regime. We find a correlation coefficient of −0.75 for the spatially-resolved data.

As a sanity test, we also estimated the log(N/O) from the N2S2 calibration (Pérez-Montero & Contini 2009), and the metallicity from the N2 parameter (Pettini & Pagel 2004), and studied the relation. Since calibrations for N/O and O/H based on nitrogen lines (such as N2S2 and N2, respectively) in principle assume a positive relation between N/O and metallicity, we expected to see a change in the trend found using the direct measurements. For example, from Pérez-Montero & Contini (2009), their Figures 5 & 6 show the existence of a strong correlation between the metallicity derived from the N2 parameter and the N/O ratio in such a way that the metallicity predicted by N2 is overestimated in objects with a high N/O. However, we find that the apparent anticorrelation between N/O and O/H in Figure 18 still holds, thereby adding confidence to our results. However the slope (of the negative trend) in this case is steeper than that seen in the relation obtained using HII-CHI-mistry. This might be due to the intrinsic increasing trend of N/O with metallicity in the secondary regime, which will pull the slope to show a positive slope, and hence will result in a steeper negative trend. In Figure 19, we only show the spatially-resolved relation based on quantities obtained from the HII-CHI-mistry code, rather than from indirect methods (N2 and N2S2), because the former is based on models which do not assume any relation between the metallicity and log(N/O). Moreover, the calibration uncertainty on indirect methods are very high (0.18 dex for (O/H)[N2] and 0.1 dex for (N/O)[N2S2] at 68% confidence level), compared to HII-CHI-mistry results. Given the spaxels are 0.1″, one may also argue about the effects of seeing. However seeing will only smear out any pre-existing trend, rather than producing a non-existent trend. The slope of the spatially-resolved data appears to be flatter than the slope of integrated data of the four H II regions. This is likely due to the limitations of the photoionisation models used in HII-CHI-mistry as described in Pérez-Montero (2014). High S/N data allowing the detection of the weak auroral line would help circumvent this problem. From the S/N of the faintest H II region (region 4), we estimate that we would need to stack around 20 spectra in order to achieve a S/N ([O III] 4363) > 3.0 within this region. However, our tests on attempting to map this auroral line in the gaps between the H II regions were not successful due to lack of S/N, even after integrating over 100s of spaxels.

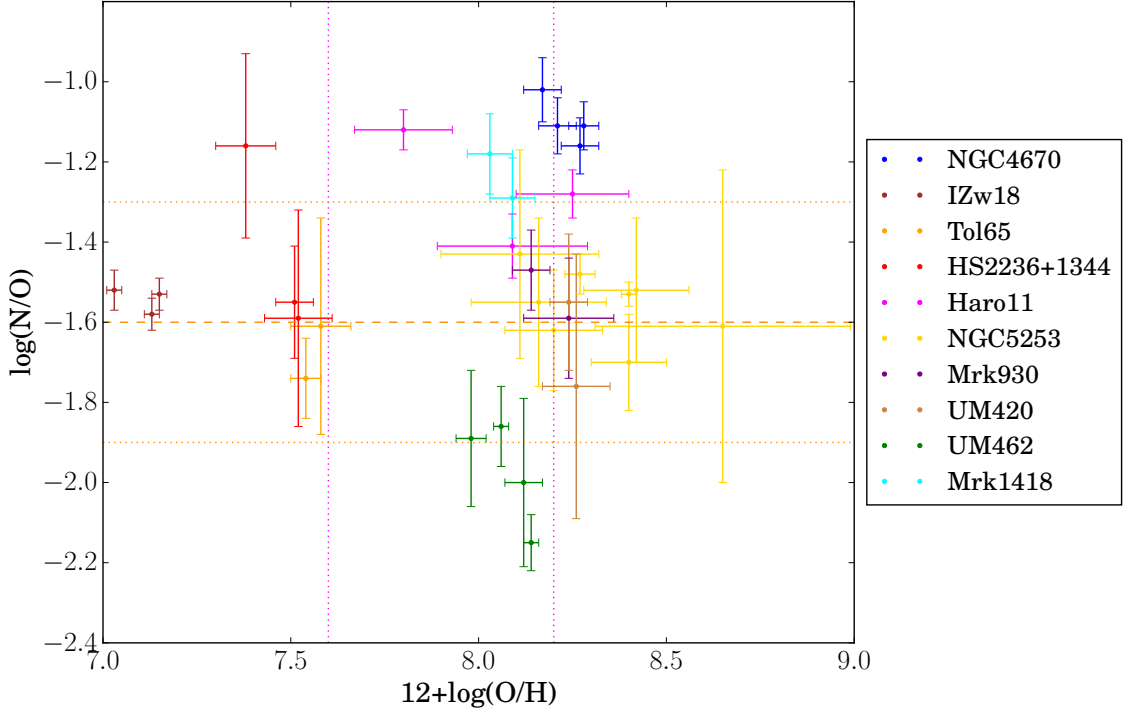

**Figure 20.** log(N/O) versus 12 + log(O/H) of H II regions within BCDs, where data points in the same colour correspond to the H II regions from the same BCD. See Table 5 for the data reference. The orange dashed line indicates the plateau at log(N/O) = −1.6, and the orange dotted lines correspond to ±0.3 dex enclosing the region of scatter found in the literature (e.g. Izotov & Thuan 1999). The vertical pink dotted lines at metallicity 12 + log(O/H) = 7.6, define the transition from low- to intermediate-metallicity regime and the transition from the intermediate-metallicity regime to the high metallicity regime at 12 + log(O/H) = 8.2.

### 4.2 H II regions in other BCDs

We further explore the negative trend observed above by studying the relation for H II regions within other BCDs. Note here that all the observational studies mentioned in the beginning of section 4.1, are based on the H II regions in different low-metallicity star-forming galaxies. None of those analyses involves the study of the relation at spatially-resolved scale in a single galaxy. Figure 20 shows this relation for star-forming regions in ten BCDs, where data points in same colour correspond to the star-forming regions in the same BCD. We have compiled the data based on the following criteria: (1) All regional values are from the integrated spectra obtained from the IFS observations. (2) The regions are selected on the basis of H$\alpha$ maps of BCDs. (3) The values of log(N/O) and 12 + log(O/H) are measured using the direct $T_e$-method. In Table 5 we specify a reference for each dataset. Two separate IFU studies were performed for NGC 5253, but we analyse data from Westmoquette et al. (2013), which includes the star-forming regions studied in Monreal-Ibero et al. (2012).

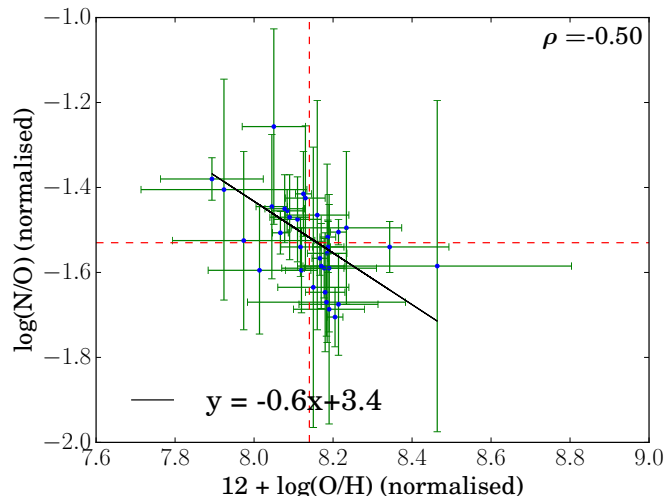

**Figure 21.** Normalised log(N/O) versus 12 + log(O/H). Error bars are the same as on the original data points (as in Figure 20). Best-fit: y= (-0.6±0.16)x+(3.4±1.3). The red dashed straight lines correspond to the medians of the log(N/O) and 12 + log(O/H).

In the sample of ten galaxies, we first analyse six galaxies individually for which there are more than two data points. The values of the Pearson correlation coefficient $\rho$ (Table C1) suggest a negative correlation for all the six galaxies. Taking error bars into account, we also fitted a constant model with and without any slope to perform a likelihood ratio test. In Appendix C, we present the values of $\rho$ and the details of the log likelihood ratio test for six galaxies in the sample. Among the six galaxies, IZw 18 and HS2236+1344 lie in the low-metallicity range where we would not expect a trend, which is also shown by the likelihood ratio test. For NGC 4670, $\rho = -0.83$, which suggests a strong anticorrelation. However the likelihood ratio test does not favour a negative trend for the group of H II regions under study. Out of the remaining three galaxies, the likelihood ratio test shows that two BCDs in the sample Haro 11 and UM 462 exhibits a negative trend.

The above analysis can not be performed on the remaining four galaxies (Mrk1418, Mrk930, UM420 and Tol 65) which have only two data points. However for one of the two regions in Tol 65, Lagos et al. (2016) report high spatially-resolved log(N/O) values, preferentially in spaxels with lower oxygen abundance, which indicates a negative trend on spaxel-by-spaxel basis.

Though in the above analysis we have found hints of a negative trend in the log(N/O) versus log(O/H), it is difficult to say anything definite about the trend because of the few data points in each galaxy. To circumvent this problem, we put the data points corresponding to each galaxy on the same scale of metallicity and log(N/O) by normalising them with the medians of the metallicities and log(N/O) for all H II regions in the ten BCDs. Figure 21 shows the relation after normalisation. We find a negative correlation coefficient of $\rho = -0.50$. While the trend was not at all clear from the scatter plot of Figure 20, there is a strong hint of a negative trend in Figure 21. The likelihood ratio test also confirms the negative trend (see appendix C). The negative slope (−0.6±0.16) which we find here agrees within errors with the slope (−0.798±0.350) found for NGC 4214 (Kobulnicky & Skillman 1996). This shows that it is quite possible that a negative trend between log(N/O) and metallicity might exist for H II regions within BCDs.

**Table 5.** References and Distances of the BCDs shown in Figure 20

| BCDs | References | Distance (Mpc) |
|---|---|---|
| NGC 4670 | This study | 18.6 |
| IZw18 | Kehrig et al. (2016) | 18.2 |
| Tol 65 | Lagos et al. (2016) | 42.7 |
| HS2236+1344 | Lagos et al. (2014) | 79.7 |
| Haro11 | James et al. (2013b) | 83.6 |
| NGC 5253 | Westmoquette et al. (2013) | 3.8 |
| Mrk930 | Pérez-Montero et al. (2011) | 83.2 |
| UM420 | James et al. (2010) | 23.8 |
| UM462 | James et al. (2010) | 14.4 |
| Mrk1418 | Cairós et al. (2009) | 14.6 |

### 4.3 Possible causes for a negative relation between N/O and O/H

In this section, we discuss possible explanations for a negative trend between metallicity and log(N/O) in spatially resolved data. In order to understand the negative trend, we need to put our spatially-resolved data of NGC 4670 in a more global perspective. Figure 22 shows our spatially-resolved data from this study (brown points) along with the global data (grey points) of a galaxy sample drawn from the MPA-JHU catalog of SDSS Data Release 7 including only objects classified as purely star-forming. They lie in the redshift range of 0.023 < z < 0.1 and their stellar masses are restricted to $7 < log(M_\star/M_\odot) \leq 9$ because BCDs are dwarfs with a stellar mass $M_\star < 10^9$ $M_\odot$ (Lian et al. 2016). We compute the log(N/O) and log(O/H) for SDSS galaxies using the HII-CHI-mistry code. In Figure 22, we find that the higher metallicity and lower (N/O) end overlaps with the normal star-forming galaxies and therefore is experiencing the typical gas mixing observed within this metallicity range. The deviation occurs for the metallicity range 12+log(O/H) $\lesssim 8.2$.

**Relation to star-formation activity and localised N enrichment:** To explain the deviation observed in Figure 22, we show in Figure 23 the spatially-resolved relation between log(N/O) versus 12 + log(O/H), obtained from the HII-CHI-mistry as explained in Section 3.4.2. The data points are color-coded with respect to the intrinsic H$\alpha$ fluxes in the corresponding spaxels. Since H$\alpha$ traces the current (< 5 Myr) star-formation activity, we use the intrinsic H$\alpha$ flux as a proxy for the on-going star-formation activity in the region of interest. We did not utilise the spatially-resolved star-formation rate maps created in Section 3.5.3 because the SFR recipes used to create these maps assume a certain metallicity (solar or sub-solar). Though the metallicity-dependence will only act as an additive constant in Figure 23, we prefer avoiding any circularity of argument in the analysis. In Figure 23, we find that the spaxels with higher metallicity and lower log(N/O) shows a lower intrinsic H$\alpha$ flux whereas the spaxels with lower metallicity and higher log(N/O) shows a higher intrinsic H$\alpha$ flux.

At the low metallicity end, the trend between low-metallicity and increased star-formation can be explained by a scenario in which metal-poor gas flows in, increasing the star-formation rate of the region and decreasing the metallicity. The velocity map of the FOV (Figure 7, left panel) shows a rotating structure, with no shock or turbulence which are typical of inflows/outflows. However gas flows have been proposed/studied in galaxies with indication of rotation (Queyrel et al. 2012; Sánchez Almeida et al. 2014, 2015; Olmo-García et al. 2017), which may be the case in NGC 4670. A possible cause of the high log(N/O) is a localised nitrogen enrichment of the interstellar medium from the ejecta of WR stars (Kobulnicky et al.

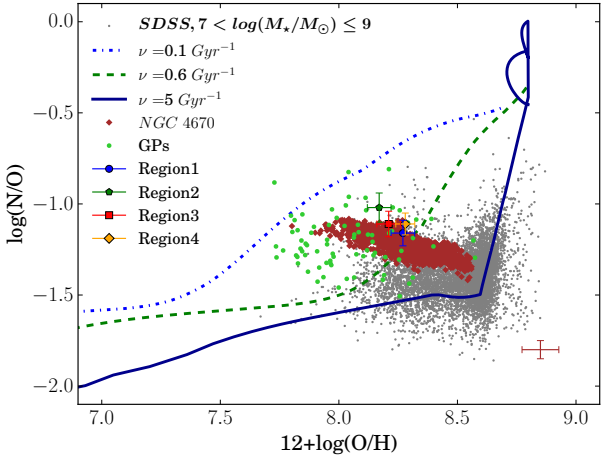

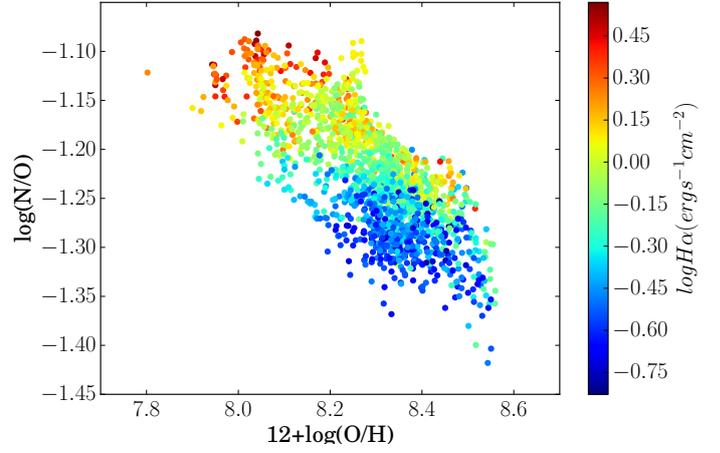

**Figure 22.** Chemical evolution models from Vincenzo et al. (2016) on the log(N/O) versus log(O/H) plane. The three curves correspond to the three different SFEs, i.e. $\nu = 5$ Gyr$^{-1}$ (solid blue), $\nu = 0.6$ Gyr$^{-1}$ (dashed green) and $\nu = 0.1$ Gyr$^{-1}$. These models are generated at the infall mass of $10^9$ M$_\odot$, and Kroupa IMF. The data corresponding to the integrated spectra of the four star-forming regions are obtained from the direct $T_e$-method and are colour-coded as in the previous figures, region1: blue, region2: green, region3: red and region4: orange. The spatially-resolved data from this work are shown as brown points with the median error bars at the right-hand corner of the plot. The SDSS DR7 data for a sample of nearby star-forming galaxies are shown by grey dots, and the data for green peas are shown by light green points. The HII-CHI-mistry code has been used to generate the values of log(N/O) and log(O/H) for spatially-resolved data from this work, SDSS galaxies and the green peas.

**Figure 23.** Spatially-resolved relationship between log(N/O) and 12 + log(O/H), where data points are obtained from HII-CHI-mistry code and are colour-coded with respect to the intrinsic H$\alpha$ fluxes in the corresponding spaxels. We find that the spaxels with lower metallicity and higher log(N/O) also show higher H$\alpha$ fluxes, whereas the spaxels with higher metallicity and lower log(N/O) show lower H$\alpha$ fluxes. H$\alpha$ being a proxy for current ($<$ 5 Myr) star-formation, the negative trend in the log(N/O) versus metallicity is apparently dependent on the on-going star-formation.

1997; Brinchmann et al. 2008; López-Sánchez & Esteban 2010). In support of this enrichment scenario, the log(N/O) map (Figure 11) also shows an increase in the region approximately coincident with the peak in WR emission (Figure 15). Note here that if metal-poor gas were inflowing, the level of nitrogen enrichment needed to produce an elevated log(N/O) for a given low log(O/H) would be significant. Note here that the inflow does not change N/O at all, only changes O/H. Even if the inflow dilutes the metallicity by say 0.2 dex, the N/O is still larger than the spaxels with that metallcity. Hence, pollution seems certainly more likely than inflow in this case because the N/O enhancement seems local.

**Varying star-formation efficiency in different star-forming regions:** In Figure 22, we also show the chemical evolution models (three curves) taken from Vincenzo et al. (2016). Though these models have been originally tested on a global basis for galaxies (SDSS sample and low-metallicity dwarfs), they can be used for our spatially-resolved data. This is because in these models, it is assumed that a galaxy is made up of a single zone within which the chemical elements mix instantaneously and uniformly. This assumption is valid for individual star-forming regions as well. Note here that no chemical evolution models are available currently for studies inside galaxies at spatially-resolved scales, hence these models provide the best alternative for the spatially-resolved study presented here. Details on these models can be found in Vincenzo et al. (2016); Matteucci (2012) including their specific parameters such as infall mass ($M_{inf}$), infall time-scale, star-formation efficiency (SFE denoted by $\nu$), IMF, mass loading factor, differential galactic outflow.

While employing the chemical evolution models, we have assumed the reference values for all parameters given in Vincenzo et al. (2016), except $M_{inf}$ and SFE. Since the BCDs are low-mass galaxies, we have assumed $M_{inf} = 10^9$ M$_\odot$. At the infall time scale of $\tau_{inf} = 0.1$ Gyr and Kroupa IMF (Kroupa et al. 1993), we vary SFEs from 0.1 Gyr$^{-1}$ to 5 Gyr$^{-1}$, shown by three curves with the purpose of covering the region on the plot occupied by the spatially-resolved data of NGC 4670. We find that the spaxels with higher log(N/O) and lower metallicity lie on the region in the plot covered by the chemical evolution model with a relatively low SFE, i.e. $\nu \sim$ 0.1 Gyr$^{-1}$. However the spaxels with lower log(N/O) and higher metallicity lie in that region which is covered by the models with a relatively high SFE, i.e. $\nu \sim 5$ Gyr$^{-1}$. The curve corresponding to an intermediate value of SFE appears to go through the middle of the distribution of (N/O)-(O/H) values. Hence we find a gradient in SFE while going from one end (high (N/O)-low (O/H)) to the other end (low (N/O)-high (O/H)) of (N/O)-(O/H) plane. A higher SFE will result in an increased production of oxygen per unit time by massive stars increasing the metallicity of the corresponding regions. If there is a delay in the production of nitrogen, the ratio of nitrogen-to-oxygen will decrease. The same argument related to the delayed production of nitrogen can be applied to the regions with lower SFE, low metallicity and high nitrogen-to-oxygen ratio. Hence it appears that the SFE at small spatial scales plays an important role in maintaining the oxygen abundance and the nitrogen-to-oxygen ratio within galaxies. It is likely through its effect on the star-formation history and on the balance of the different stellar population which contribute to the oxygen and nitrogen abundance (Vincenzo et al. 2016).

However this scenario does not tie in easily with the increased H$\alpha$ at high log(N/O) and low metallicity. An increased H$\alpha$ indicates a higher SFR, while the same regions are covered by lower SFE as seen in Figure 22. This indicates a higher star-formation rate in a region with lower SFE. If we define SFE as the ratio of SFR to the available gas for forming stars, then a high SFR for a region with lower SFE simply implies that a large amount of gas is available in the region which is not forming stars. To explore this hypothesis, a more complete analysis of the spatial distribution of the relative abundance of neutral, molecular and ionised gas, are required. High SFR with low SFE may also point towards the fact that star-formation is a self-regulating process at local sub-galactic scales. Regions with high SFRs may have strong negative energetic feedback from the massive dying stars, which will result in a decreased SFE in the respective star-forming regions.

**Alternative explanation:** In Figure 22, we also show data corresponding to green pea galaxies which are low-redshift extreme emission-line galaxies with very high specific SFR, that are also thought to be excellent analogs of high-redshift star-forming galaxies (Cardamone et al. 2009). As we can see in Figure 22, they exhibit relatively high log(N/O) at a given metallicity in the low-to-intermediate metallicity regime compared to the SDSS galaxies. As discussed in Amorín et al. (2010, 2012a), the possible cause of such a behaviour could be the simultaneous inflow of metal-poor gas induced by interactions and the outflow of metal-rich gas driven by supernova remnants and stellar winds. The scenario involving gas dynamics was proposed by other observational (van Zee et al. 1998) and theoretical studies (Köppen & Hensler 2005) as well. In our study, we have found a relative increase in the He abundance in the region of NGC 4670 surrounding the region and the WR bump, which suggest the local effect of a strong wind emanating from the WR stars. The age of the stellar population found in our work is 10–30 Myr which are old enough for the existence of supernovae remnants and stellar winds from them. Further analysis is required to find the dynamical state of the ionised, molecular and neutral gas in this galaxy.

## 5 SUMMARY

We have carried out a spatially-resolved study of four H II regions and the surrounding ionised gas located in the central region of the BCD NGC 4670 using the GMOS-N IFS data at the spatial scale of 9 pc. A summary of our main results is as follows:

(i) The radial velocity map indicates a slow rotation of the ionised gas varying between $\sim -10$ to 30 km s$^{-1}$ about an isovelocity S-shaped contour. The velocity dispersion varies between 40–104 km s$^{-1}$, with relatively lower dispersion in the H II regions. No signatures of shock or complex velocity structure is found. The classical [S II]/H$\alpha$ and [N II]/H$\alpha$ emission line ratio or BPT diagrams show that photoionisation by massive stars is the main source of ionisation.

(ii) In the integrated spectra of the four H II regions we could detect the auroral lines [O III] $\lambda$ 4363, [O II] $\lambda\lambda$ 7320, 7330 which we use to estimate T$_e$ ([O III]), N$_e$ ([S II]) and T$_e$ ([O II]), and hence calculate 12 + log(O/H) varying between 8.17–8.28, log(N/O) varying between $-1.02$ to $-1.16$ and the helium abundance y$^+$(5876) varying between 0.086–0.093 for the four H II regions.

(iii) In the spatially-resolved data of the FOV, the auroral line [O III] $\lambda$ 4363 could not be detected with enough S/N (>3), which prevented the chemical abundance determination using the direct T$_e$-method. Hence we use the HII-CHI-mistry (v3.0) to map 12 + log(O/H) and log(N/O), which vary between 7.80–8.56 and between $-1.4 - -1.08$ respectively across the FOV. We then use the 12+log(O/H) map to create T$_e$ and N$_e$ maps, with which we map helium abundance y$^+$(5876) varying between 0.05–0.12.

(iv) The age map is created at a metallicity of 0.35Z$_\odot$ by comparing the observed EW(H$\alpha$) map with the modelled EW map using the STARBURST99 population synthesis model. The resultant age map varies between 10–30 Myr.

(v) We also estimate the number of WR stars from the integrated spectrum of the WR region, which overlaps with three of four H II regions. The WR region contains approximately $\sim$125$\pm$4 WC4 stars, $\sim$60$\pm$20 WN7-9 stars and $\sim$480$\pm$50 WN2-4 stars. The estimate is done by using the LMC WR templates from Crowther & Hadfield (2006).

(vi) We map SFR of the FOV from the dereddened H$\alpha$ luminosity at sub-solar metallicity of 0.35 Z$_\odot$ which varies between 0.02–0.49 $\times$ 10$^{-3}$ M$_\odot$ yr$^{-1}$.

(vii) We revisit the N/O conundrum by studying the relation between the spatially-resolved 12 + log(O/H) and log(N/O) and the integrated data of the H II regions from ten other BCDs from the literature, and found an unexpected negative trend. We also compare the spatially-resolved data with the low-mass SDSS galaxies and the green pea galaxies and explore various scenarios to explain the trend including nitrogen enrichment, and variations in star formation efficiency via chemical evolution models.

In summary, the present analysis shows that a negative trend between log(N/O) and 12+log(O/H) may exist at the spatially-resolved scales and H II regions within BCDs. It is possible that the negative trend which we observe in our spatially-resolved data is merely a manifestation of the large scatter which has been observed previously in other low-metallicity galaxies. To confirm the behaviour of log(N/O) with metallicity, these two properties and their relation need to be studied with high S/N data at spatially-resolved scales across a large sample of BCDs. Those datasets would allow us to map the distribution of metallicity and log(N/O) from robust direct methods and also analyse the WR population in the entire galaxy for a plausible effect of chemical pollution. Moreover such datasets would enable the study of variation of the above-mentioned observables not only within galaxies but also within H II regions, for example via radial profile analysis or a segmentation analysis of individual H II regions (López-Hernández et al. 2013). Those studies would be possible in the future with the current and upcoming IFS facilities like MUSE on Very Large Telescope, WEAVE on William Herschel Telescope and Keck Cosmic Web Imager because of their large FOV and good spatial sampling, which will enable us to map chemical abundances of entire galaxies. Moreover, such studies could also investigate kinematic signatures to further probe the possible cause of the observations related to gas dynamics. As we have shown here, the combination of chemical and kinematical analyses is essential in disentangling chemical evolution scenarios, and obtaining such information across a large sample of entire systems will be key in understanding the origin of nitrogen in these less chemically evolved star-forming systems. In so doing, we will enhance our understanding of chemical evolution scenarios in galaxies within the high-redshift Universe.

## ACKNOWLEDGEMENTS

It is a pleasure to thank Jose Vilchez for various useful discussions during the course of this work. We also thank Alessandra Aloisi, Tracy Beck and Mark Westmoquette for their invaluable help during the early stages of this project. NK thanks Fiorenzo Vincenzo for providing with the chemical evolution models. We thank the referee for a positive feedback. NK thanks the Institute of Astronomy, Cambridge and the Nehru Trust for Cambridge University for the financial support during her PhD. BLJ thanks support from the European Space Agency (ESA). RA acknowledges the support from the ERC Advanced Grant 695671 QUENCH. EPM thanks financial support from Spanish MINECO project AYA2016-797124-C4-4-P and CSIC grant EPM461-201650I042 and Kavli Institute for Cosmology in Cambridge for kindly hosting him in July 2017 when part of this work was made. This research made use of the NASA/IPAC Extragalactic Database (NED) which is operated by the Jet Propulsion Laboratory, California Institute of Technology, under contract with the National Aeronautics and Space Administration; SAOImage DS9, developed by Smithsonian Astrophysical Observatory"; Astropy, a community-developed core Python package for Astronomy (Astropy Collaboration et al. 2013).

## APPENDIX A: FORMULAE

We present here some of the formulae used in this work, which are actually taken from Pérez-Montero (2017), but have some misprints in the published version. The correct expressions are as follows:

- Temperature of low-ionisation zone:

$$t([O\textsc{ii}])(10^4 K) = a_0(n) + a_1(n)R_{O2} + \frac{a_2(n)}{R_{O2}} \quad (A1)$$

where

$$R_{O2} = \frac{I(3726) + I(3729)}{I(7319) + I(7330)} \quad (A2)$$

$$a_0(n) = 0.2526 - 0.000357n - \frac{0.43}{n} \quad (A3)$$

$$a_1(n) = 0.00136 + 5.431 \times 10^{-6}n + \frac{0.00481}{n} \quad (A4)$$

$$a_2(n) = 35.624 - 0.0172n + \frac{25.12}{n} \quad (A5)$$

- The ratio $N^+/O^+$

$$\begin{aligned}log(\frac{N^+}{O^+}) = &\; log(\frac{I(6583)}{I(3726) + I(3729)}) + 0.493 - 0.025t_l \\ &- \frac{0.687}{t_l} + 0.1621 log(t_l)\end{aligned} \quad (A6)$$

where $t_l$ is the low-ionisation zone temperature in units of $10^4 K$.

## APPENDIX B: ELECTRON TEMPERATURE AND DENSITY DERIVED FROM METALLICITY

Figure B1 (upper panel) shows the $T_e$([O III]) map obtained from the metallicity map (Figure 10), which was itself derived from the HII-CHI-mistry code. Using this indirectly obtained $T_e$ map and the [S II] doublet ratio, we derived the $N_e$ map (Figure B1, lower panel). We find that the majority of spaxels show 50 cm$^{-3}$, which is actually an upper-limit on the electron density. This shows that the region under study is low-density region.

## APPENDIX C: ANALYSIS OF LOG(N/O) VS 12 + LOG(O/H) FOR HII REGIONS WITHIN BCDS

The correlation-coefficient for the six galaxies are given in Table C1.

We performed the likelihood ratio test (Neyman-Pearson test) for six out of ten BCDs to find out if there exists a negative trend between log(N/O) and log(O/H) for H II regions within BCDs. For each BCD, we fit two straight lines of the form: y = c and y = mx+c, where m is the negative slope and c is the constant. The individual straight line fits for the six galaxies are shown in Figure C1. All these fits take into account the error bars on the fitted data. The corresponding $\chi^2$ values are tabulated in Table C2. $\chi^2_{y=c} - \chi^2_{y=mx+c}$ is known to follow a $\chi^2_1$ distribution. Since for a $\chi^2$ distribution, $variance(\chi^2) = \sigma^2 = 2\nu$, hence we have $\sigma = \sqrt{2}$. The results of the test are shown in Table C2. We find that a straight line of the form y = c, fits four out of six galaxies which includes NGC 4670. Only two galaxies Haro 11 and UM462 show a negative trend. The result of the likelihood ratio test on the normalised data for all H II regions (Figure 21) are shown in the last column, which show that a negative correlation exits between log(N/O) and log(O/H).

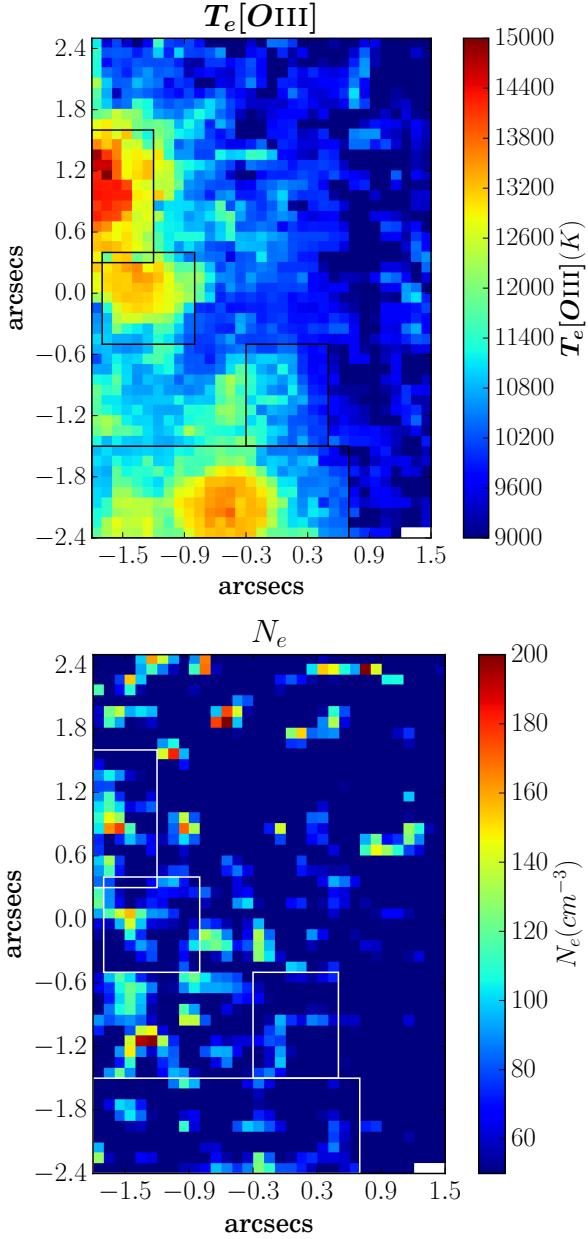

**Figure B1.** Upper panel: $T_e$([O III]) map obtained from the metallicity map (Figure 10). Lower panel: $N_e$ map using the [S II] doublet ratio map and $T_e$([O III]) map on left. Majority of the spaxels have upper-density limit of 50 cm$^{-3}$ in those spaxels. The four rectangular boxes in the two maps show the location of the four H II regions.

**Table C1.** Pearson correlation coefficient for individual galaxies.

| Galaxy | $\rho$ |
|---|---|
| NGC4670 | -0.83 |
| IZw18 | -0.50 |
| HS2236+1344 | -1.00 |
| Haro11 | -0.68 |
| UM462 | -0.79 |
| NGC5253 | -0.45 |

**Table C2.** Likelihood ratio test fit for individual galaxies.

| Galaxy | $\chi^2_{y=c}$ | $\chi^2_{y=mx+c}$ | $\chi^2_{y=c} - \chi^2_{y=mx+c}$ | $\frac{\chi^2_{y=c} - \chi^2_{y=mx+c}}{\sigma}$ |
|---|---|---|---|---|
| NGC4670 | 1.758 | 0.713 | 1.045 | 0.7 |
| IZw18 | 1.152 | 0.961 | 0.191 | 0.1 |
| HS2236+1344 | 2.340 | 0.001 | 2.339 | 1.7 |
| Haro11 | 10.592 | 5.381 | 5.211 | 3.7 |
| UM462 | 6.517 | 1.560 | 4.957 | 3.5 |
| NGC5253 | 3.554 | 3.122 | 0.432 | 0.3 |
| 10 galaxies | 31.896 | 23.085 | 8.811 | 6.2 |

This paper has been typeset from a T<sub>E</sub>X/L<sup>A</sup>T<sub>E</sub>X file prepared by the author.

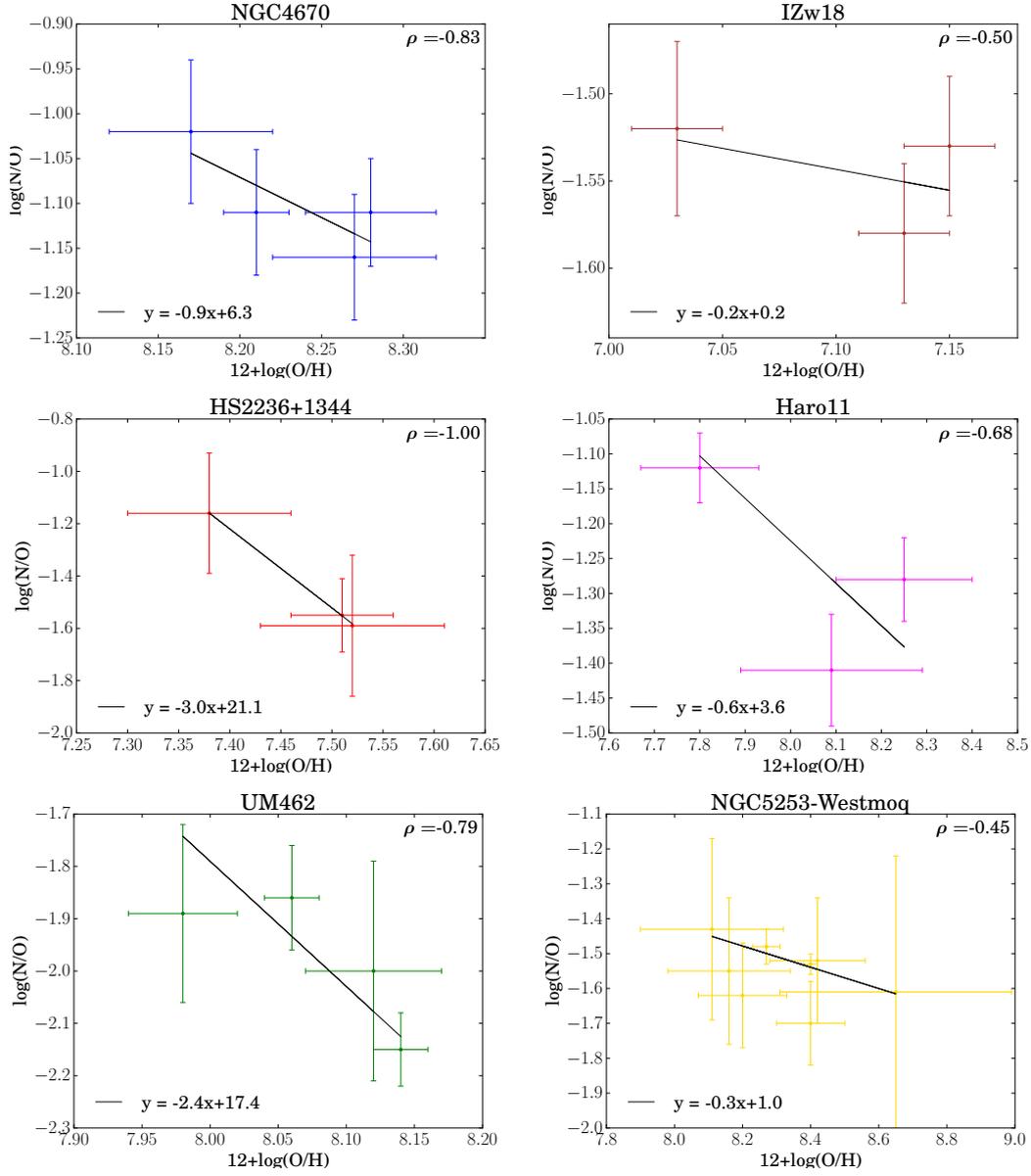

**Figure C1.** The individual straight line fits for the six galaxies, which have more than 2 data points. The best-fit parameters is shown at the bottom-left corner in each panel, where y denotes log(N/O) and x denotes 12 + log(O/H). The Pearson correlation co-efficient is shown in the top-right corner of each panel, which is negative for six galaxies.